\documentclass{ccjnl}
\pdfoutput=1

\makeatletter
\def\maketitle{
  \twocolumn[
  \begin{@twocolumnfalse}
    {\tikz\fill[fill=white] (0,0) rectangle (\textwidth,0.3);
    \vskip-11pt
    {\parbox[t]{\boxwidth}{\vfill\hskip-3pt\parbox{\boxwidth}{\raggedright\large\sffamily\scshape\bfseries\journal}\vfill}}\vskip1cm
    \raggedright{\LARGE\sffamily\bfseries\@title\par}\vskip12pt
    \titlerule\\
    {\bfseries\fontsize{10.5pt}{12pt}\selectfont\@author}\vskip12pt
    \scriptsize\CCJNLaddress\CCJNLcorinfo
    \vskip1cm}
  \end{@twocolumnfalse}
  ]
}
\makeatother

\usepackage{stfloats}
\title{RIS-Aided Cooperative ISAC Network for Imaging-Based Low-Altitude Surveillance}
\author{Zhixin Chen\inst{1}, Yixuan Huang\inst{2}, Zhengze Ji\inst{1}, Jie Yang\inst{3,4,*} and Shi Jin\inst{2,4,*}\corinfo{yangjie@seu.edu.cn, jinshi@seu.edu.cn}}

\address[1]{Chien-Shiung Wu College, Southeast University, Nanjing, China}
\address[2]{National Mobile Communications Research Laboratory, Southeast University, Nanjing, China}
\address[3]{Key Laboratory of Measurement and Control of Complex Systems of Engineering, Ministry of Education, Southeast University, Nanjing, China}
\address[4]{Frontiers Science Center for Mobile Information Communication and Security, Southeast University, Nanjing, China}

\usepackage{xcolor}
\begin{document}
\maketitle

\begin{abstract}
The low-altitude economy is integral to the advancement of numerous sectors, necessitating the development of advanced low-altitude surveillance techniques. 
Nevertheless, conventional methods encounter limitations of high deployment costs and low signal strength. 
This study proposes a reconfigurable intelligent surface (RIS)-aided cooperative integrated sensing and communication (ISAC) network for low-altitude surveillance. 
This network employs RISs to reflect ISAC signals into low-altitude space for sensing. 
To enhance signal strength, we employ active RIS (ARIS) to amplify the signals. 
Moreover, in order to avoid error propagation and data association in traditional sensing methods, we model low-altitude surveillance as an imaging problem based on compressed sensing theory, which can be solved through the subspace pursuit algorithm. 
We derive the Cram\'{e}r-Rao lower bound (CRLB) of the proposed RIS-aided low-altitude imaging system and analyze the impacts of various system parameters on sensing performance, providing guidance for ISAC system configuration. 
Numerical results show that ARIS outperforms passive RIS under identical power constraints, achieving effective imaging and target detection at altitudes up to 300 meters. 
\keywords{Low-altitude wireless network, integrated sensing and communications, wireless imaging, reconfigurable intelligent surface, compressed sensing, Cram\'{e}r-Rao lower bound}
\end{abstract}

\section{Introduction}
\label{s1}

As a pivotal strategic domain for future airspace resource utilization, low-altitude economy (LAE) has received widespread attention in recent years, with research focusing on low-altitude wireless networks (LAWNs) \cite{luo2025toward,how_to_achieve,11059622}. 
These networks primarily focus on providing communication and sensing capabilities below 300 meters, facilitating essential services such as logistics, surveillance, and air mobility \cite{9631203}. 
Low-altitude surveillance is crucial for the development of LAE, as it ensures the safe integration of unmanned aerial vehicles (UAVs), birds, and other airborne objects into the aerial space, and prevents illegal activities such as spying, hacking, and target attacks \cite{khan2022detection}. 
To meet these demands, LAWNs must offer persistent spatiotemporal environmental sensing and communication \cite{luo2025toward}.

Low-altitude surveillance is a challenging task due to the small size of airborne objects such as UAVs and birds \cite{anti-drone}. 
Video systems, such as optical cameras and infrared sensors, are traditionally employed for monitoring low-altitude airspace \cite{wu2024vehicle}. 
However, their performance is heavily dependent on weather conditions, lighting, and line-of-sight observation \cite{khan2022detection}. 
Radar is renowned for its high target detection and localization accuracy \cite{anti-drone}. 
However, radars involve high deployment costs for large-scale networks \cite{learned_off_grid}, and their transmit power is restricted when deployed in residential areas \cite{how_to_achieve}.

Low-altitude surveillance supports the UAV services and intrusion detection applications, both of which are agreed-upon use cases of the 3rd Generation Partnership Project (3GPP) \cite{li2023integrated}. 
Integrated sensing and communication (ISAC) leverages pure communication signals that do not depend on lighting conditions. It has the potential to address the high-cost issue by enabling simultaneous wireless communication and environmental sensing through shared infrastructure and spectrum resources, without the need for extra hardware \cite{learned_off_grid}.
While ISAC systems offer substantial advantages, the significant spatial separation between ground users and low-altitude regions poses a challenge for the seamless integration of communication and sensing. To support low-altitude surveillance, existing research typically requires specific allocation of time or beam resources for the aerial space, which may lead to communication performance degradation \cite{learned_off_grid,li2023toward,wang2024heterogeneous}.

To mitigate this degradation, one potential solution is to deploy reconfigurable intelligent surfaces (RISs) on the ground or rooftops \cite{he2024device,11008547}. 
RISs can programmatically alter the propagation properties of an incoming signal. 
By utilizing RISs, communication signals transmitted to ground users can be redirected toward the low-altitude space, thereby facilitating low-altitude surveillance without compromising the performance of the communication network \cite{he2024device}. 
Due to their low complexity and low cost, RISs are suitable for large-scale deployment \cite{wang2024reconfigurable,encinas2025riloco}. 
However, most existing RIS-aided ISAC systems utilize passive RIS (PRIS) \cite{huang2024ris,huang2023joint, zhu2023ris, li2024radio}, which face the challenge of multiplicative fading, leading to weak received signal power \cite{zhou2023framework,saikia2024ris}.

Alternatively, active RIS (ARIS) is employed to overcome the limitations associated with PRIS in this study. ARIS integrates an active reflection-type amplifier to amplify the incident signal \cite{khoshafa2021active,zhang2025research}. 
Consequently, ARIS effectively reduces the transmit power required for low-altitude surveillance. 
However, the implementation of ARIS introduces two notable drawbacks. 
First, the thermal noise introduced by ARIS elements cannot be neglected as is done for PRIS \cite{zhang2022active}. 
Second, ARIS consumes additional power consumption due to the signal amplification process \cite{khoshafa2021active}. 
Although it has been concluded that ARIS achieves superior communication performance \cite{zhang2022active}, a comparison of their imaging performance in low-altitude surveillance scenarios has not been conducted.

When it comes to sensing algorithm design, traditional ISAC-based low-altitude sensing methods, such as beamforming and localization, face several drawbacks. Beamforming is time-consuming and inefficient for the sensing of the large aerial space, while localization methods are prone to channel estimation errors in delay and angle parameters, leading to degraded sensing performance \cite{learned_off_grid}. 
Therefore, we refer to the method proposed in \cite{learned_off_grid}, which treats the entire low-altitude area as an image and models low-altitude surveillance as a compressed sensing (CS)-based imaging problem that can be solved through the subspace pursuit (SP) algorithm. 
Furthermore, we consider the employment of PRIS and ARIS, which eliminates the need for specialized beamforming design at the base stations (BSs). 
The RIS-aided system can intelligently reflect ISAC signals to the low-altitude space for sensing, thus enhancing the system's capability without the complexity of traditional beamforming techniques. 
By varying RIS phases, multiple sensing perspectives toward the low-altitude space can be achieved, resulting in numerous channel state information (CSI) measurements, which contain sufficient aerial image information. 

However, CS-based imaging algorithms lack proper theoretical frameworks for analyzing their fundamental performance limits. \cite{huang2024ris} uses the point spread function to analyze the constraints on sensing accuracy due to channel correlations, but only provides a qualitative analysis without offering a quantitative evaluation of imaging accuracy. 
\cite{huang2024fourier} derives the diffraction-limited resolution, but this metric does not account for the effects of noise. 
Given that localization algorithms typically use the Cram\'{e}r-Rao lower bound (CRLB) as a performance limit indicator \cite{huang2023joint}, we draw on the work in \cite{li2024radio} to derive the CRLB of CS-based low-altitude surveillance performance. 
Using this metric, the theoretical performance of PRIS- and ARIS-aided systems is analyzed, indicating the influence of distinct system configurations on low-altitude sensing and providing guidance for system configurations. 

The contributions of this study can be summarized as follows:
\begin{itemize}
	\item We propose a RIS-aided cooperative ISAC network for imaging-based low-altitude surveillance, employ the SP algorithm to reconstruct low-altitude images, and analyze the sensing accuracy and energy efficiency of PRIS and ARIS for low-altitude surveillance.
	\item We derive the CRLB of CS-based low-altitude imaging algorithms. An approximated version of the CRLB is derived to analyze imaging performance across different voxels and examine the impacts of transmitter (TX), receiver (RX), and RIS locations.
	\item We conduct extensive simulations to compare the imaging accuracy and detection rate of PRIS-aided and ARIS-aided networks, demonstrating the effectiveness of ARIS in sensing performance and energy efficiency. Furthermore, the conclusions derived from the CRLB-based analysis can be validated through simulations, and the SP algorithm is shown to achieve imaging accuracy approaching the CRLB under 300 meters altitude.
\end{itemize}

The remainder of this paper is organized as follows: Sec. \ref{s2} introduces the system model. Sec. \ref{s3} formulates the low-altitude surveillance problem. Sec. \ref{s4} derives the CRLB and studies the \textcolor{blue}{influence} of various system configurations, and Sec. \ref{s5} demonstrates numerical results. Sec. \ref{CONCLUSION} concludes this paper.

\textit{Notations}--Scalars (e.g., $a$) are denoted in italics, vectors (e.g., $\mathbf{a}$) in bold lowercase, and matrices (e.g., $\mathbf{A}$) in bold uppercase. The modulus of $a$ is represented as $\left | a \right | $, and the $l_2$-norm of $\mathbf{a}$ is given by $\left \| \mathbf{a}  \right \| _2$. The notation $\text{diag}\left ( \mathbf{a}  \right ) $ represents a diagonal matrix with elements from $\mathbf{a}$. $\mathbf{A}^{\text{T}}$ and $\mathbf{A}^{\text{H}}$ denote the transpose and Hermitian (conjugate transpose) of $\mathbf{A}$, respectively. $\mathbb{C} $ denotes the complex field, and $\mathbb{R} $ denotes the real field, and the imaginary unit is denoted by $j=\sqrt{-1}$.

\begin{table*}[t]
\centering
\caption{Notations of key variables.}
\label{tab1}
\begin{tabular}{p{0.06\textwidth} p{0.39\textwidth} p{0.06\textwidth} p{0.39\textwidth}}\toprule
Notation & Definition &Notation &Definition\\\midrule
$N_{\text{TX}}$&number of transmitting antennas &$N_{\text{RX}}$&number of receiving antennas\\
$T$&number of RISs&$M_{t}$&number of tunable elements in the $t$-th RIS\\
$K$&number of symbols transmitted by the TX&$N$&number of voxels in the ROI\\
$\mathbf{x}$&the scattering coefficients of all voxels in the ROI&$\mathbf{s}_k$&the $k$-th transmitted signal\\
$\mathbf{v}_k$&the received noise at the RX during the $k$-th symbol interval&$\mathbf{z}_{k,t}$&the noise introduced by the $t$-th ARIS during the $k$-th symbol interval\\
$\mathbf{\Phi }_{k,t}$&the $t$-th RIS element gains and phase shifts during the $k$-th symbol interval&$\alpha _{\text{max} }$&the maximum amplification gain of RIS\\
$h_{i,k,t,j}$&the path starting from the $i$-th transmitting antenna during the $k$-th emission, scattered by the $t$-th RIS, and ultimately reaching the $j$-th receiving antenna&$\mathbf{h}_{\text{TX}_i,\text{s}_t}$&the channel between the $i$-th transmitting antenna and the $t$-th RIS\\
$\mathbf{H}_{\text{s}_t,\text{v}}$&the channel between the $t$-th RIS and voxels&$\mathbf{h}_{\text{v},\text{RX}_j}$&the channel between voxels and the $j$-th receiving antenna\\
$P_{\text{PRIS}}$&power consumption of the PRIS-aided system&$P_{\text{ARIS}}$&power consumption of the ARIS-aided system\\
$P_{\text{TX,PRIS}}$ &the transmit power of the TX in the PRIS-aided system&$P_{\text{TX,ARIS}}$&the transmit power of the TX in the ARIS-aided system\\
$\mathbf{y}$&the CSI measurements of the imaging path&$\mathbf{n}$&the additive noise originating from channel estimation and multi-path effects\\
$L$&$L=KTN_{\text{TX}}N_{\text{RX}}$, number of CSI measurements&$\mathbf{A}$ &the sensing matrix\\
$\mathbf{u}$ &the support set of $\mathbf{x}$&$S$&sparsity value of $\mathbf{x}$\\
$\gamma$ &the reciprocal of the power of noise $\mathbf{n}$&$C_n$&the CRLB of the scattering coefficient $x_n$\\
$a$&$a = \left| \omega_{k,t,m} \right|^2$\\\bottomrule
\end{tabular}
\end{table*}

\section{System Model}
\label{s2}
\begin{figure}[t]
\centering
\subfloat[Illustration of RIS-aided cooperative ISAC network]{\includegraphics[width=0.47\textwidth]{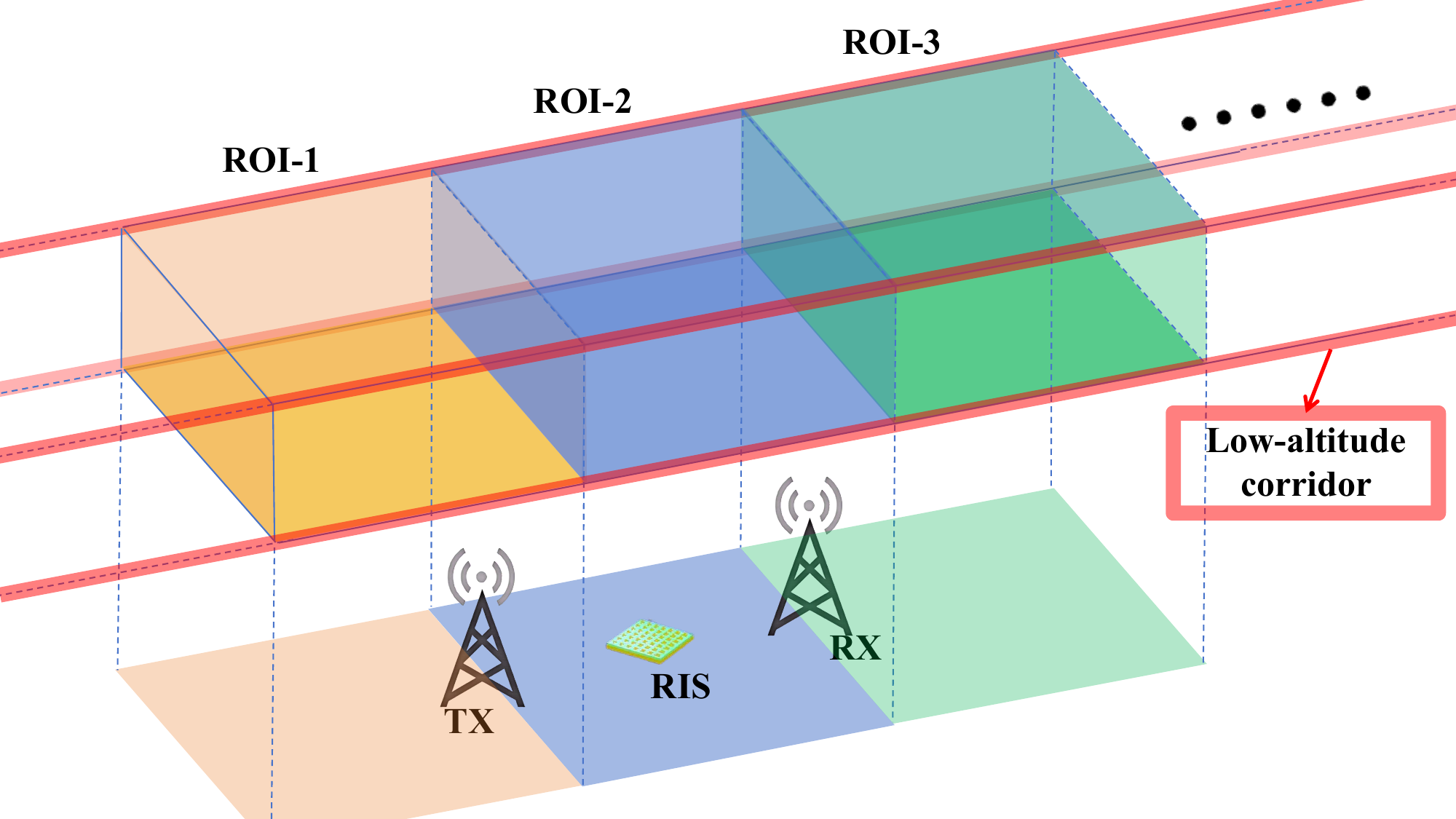}\label{f1-a}}\\
\subfloat[Illustration of the RIS-aided ISAC system for a single ROI]{\includegraphics[width=0.47\textwidth]{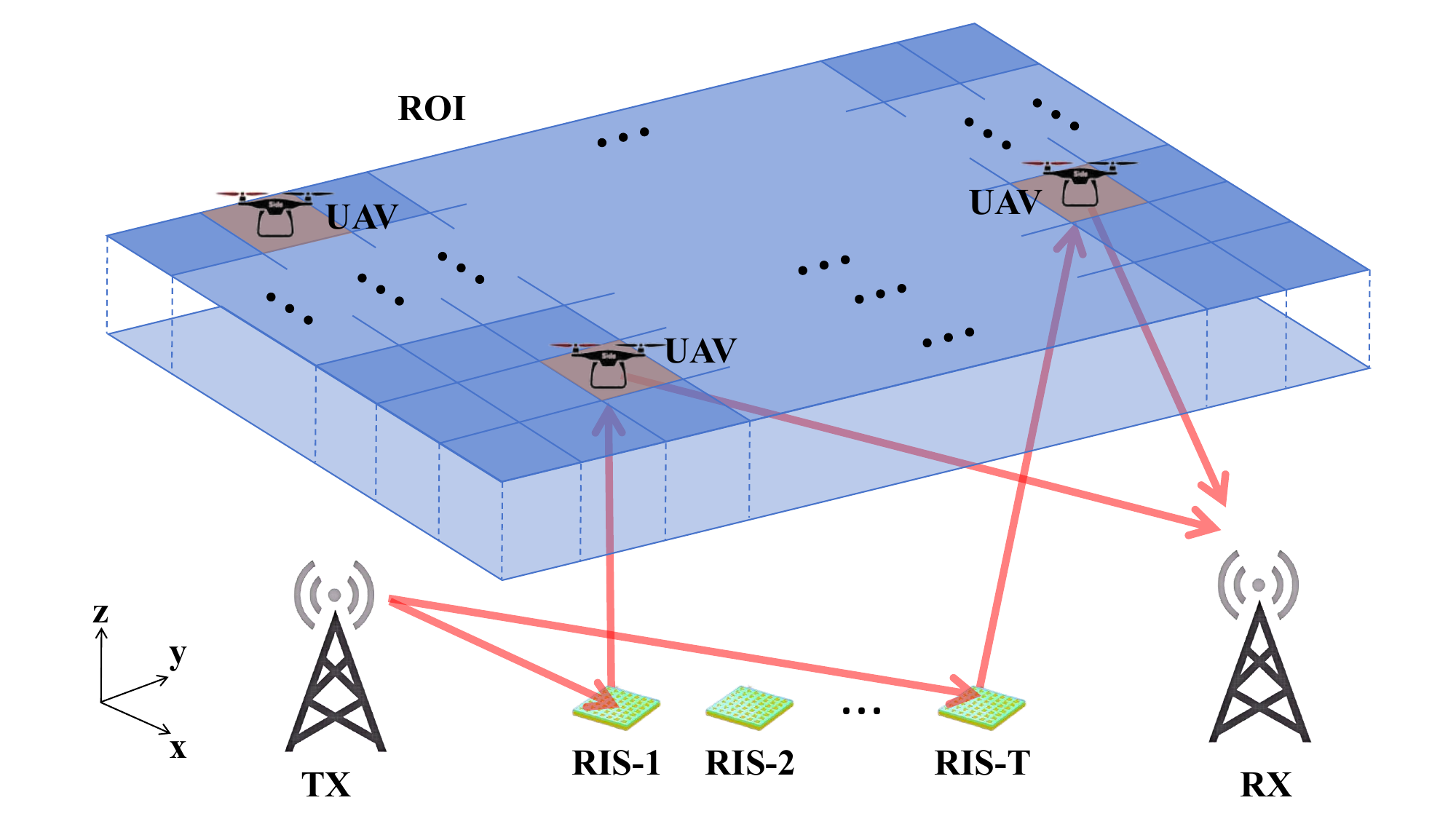}\label{f1-b}}
\caption{RIS-aided cooperative ISAC network for imaging-based low-altitude surveillance.}\label{f1}
\end{figure}

We consider a RIS-aided cooperative ISAC network operating in the three-dimensional (3D) space $\mathbb{R} ^{3} = \left \{ \left [ x,y,z \right ]^{\text{T}}:x,y,z \in \mathbb{R}   \right \}$. Due to the challenges of achieving wide-area sensing across the entire low-altitude corridor, the linear corridor is divided into several ROIs, as shown in Figure~\ref{f1-a}. This study only considers the sensing scheme design for a single ROI with the proposed RIS-aided cooperative ISAC system. The related analysis can be extrapolated to other ROIs. Beneath the considered ROI, a ground multiple-input multiple-output system with multiple RISs is employed to realize low-altitude imaging and flight activity surveillance.

As shown in Figure~\ref{f1-b}, the system includes a TX with $N_{\text{TX}}$ antennas and an RX with $N_{\text{RX}}$ antennas. Additionally, $T$ RISs are deployed horizontally to reflect the communication signals to the aerial space, enabling low-altitude imaging. The $t$-th RIS consists of $M_{t}$ tunable units. The TX transmits $K$ symbols toward the RISs via beamforming, and RIS phase configurations are altered for each symbol interval. The ROI is modeled as a 3D space and is discretized into $N=N_{\text{x}}\times  N_{\text{y}}\times N_{\text{z}}$ voxels. The scattering coefficient of the $n$-th voxel is denoted as $x_n$. $x_n$ equals 0 when the $n$-th voxel includes no targets. The scattering coefficients of all voxels form a column vector $\mathbf{x} = [x_{1}, \dots, x_{N}]^{\text{T}} \in \mathbb{R}^{N \times 1}$, which is the low-altitude image to be estimated, illustrating low-altitude targets' locations \cite{learned_off_grid}. Given that the number and size of UAVs are small compared to the vast ROI in the aerial space, the vector $\mathbf{x}$ exhibits high sparsity, i.e., most of its elements are zero.

\subsection{Signal Model}
\label{s2-1}

Let $\mathbf{s}_k=\left [ s_{k,1},\cdots ,s_{k,N_{\text{TX} }} \right ] ^\text{T} \in \mathbb{R}^{N_{\text{TX}}\times 1}$ denote the $k$-th transmitted signal, satisfying the unit norm constraint $\left \| \mathbf{s}_k \right \|_2 ^{2} = 1$. The received signal at the RX during the $k$-th symbol interval is given by
\begin{equation}
\mathbf{r} _k=\sqrt{P_{\text{TX} }}\mathbf{H}_k \mathbf{s}_k+\tilde{\mathbf{n}} _{k}   ,
\label{eq1-1}
\end{equation}
where $P_{\text{TX} }$ is the transmit power of the TX, and $\mathbf{H}_k \in \mathbb{C} ^{N_{\text{RX} }\times N_{\text{TX} }}$ denotes the channel from the TX to the RX. $\tilde{\mathbf{n}} _{k}$ is additive noise, which takes different forms for distinct types of RISs. 

In the PRIS-aided system, $\tilde{\mathbf{n}} _{k}$ only represents the received noise $\mathbf{v}_k\in \mathbb{C} ^{N_{\text{RX} }\times 1}$ at the RX. However, in the ARIS-aided system, the $t$-th ARIS introduces additional noise $\mathbf{z}_{k,t}\in \mathbb{C} ^{M_t \times 1 }$ when amplifying the incident signals. Therefore, $\tilde{\mathbf{n}} _{k}$ can be represented as \cite{chen2025experimental}
\begin{equation}
\tilde{\mathbf{n}} _{k} =\sum_{t=1}^{T}  \mathbf{H}_{\text{RX}, \text{s}_t }\mathbf{\Phi }_{k,t}\mathbf{z}_{k,t} +\mathbf{v}_k,
\label{eq1-2}
\end{equation}
where $\mathbf{H}_{\text{RX}, \text{s}_t }\in \mathbb{C} ^{N_{\text{RX} }\times M_t} $ represents the channel response between the $t$-th RIS and the RX. The diagonal matrix $\mathbf{\Phi }_{k,t}=\text{diag}  \left ( \left [ \omega _{k,t,1},\cdots ,\omega _{k,t,M_t} \right ] \right )\in \mathbb{C} ^{M_t \times M_t }$ characterizes the $t$-th RIS element gains and phase shifts. For ARISs, $1\le \left | \omega _{k,t,m} \right |^2\le \alpha _{\text{max} }$, where $\alpha _{\text{max} }$ is the maximum amplification gain, whereas $\left | \omega _{k,t,m} \right |^2  \triangleq1$ for PRISs.

According to \cite{zhou2023framework}, we have $\mathbf{z}_{k,t}\sim \mathcal{CN} \left ( \mathbf{0}_{M_t},\sigma ^2_{\text{RIS} }\mathbf{I}_{M_t}   \right ) $ and $\mathbf{v}_k\sim \mathcal{CN}\left ( \mathbf{0}_{N_{\text{RX} }}, \sigma_{\text{RX} }^2 \mathbf{I}_{N_{\text{RX} }}\right ) $, where the noise powers are $\sigma_{\text{RIS} }^2$ and $\sigma_{\text{RX} }^2$, respectively. $\mathcal{CN}\left (\boldsymbol{\mu } , \mathbf{\Sigma } \right )$ represents the distribution of a circularly symmetric complex Gaussian random vector with mean vector $\boldsymbol{\mu }$ and covariance matrix $\mathbf{\Sigma }$. $\mathbf{0}_l$ represents a zero vector of length $l$, and $\mathbf{I}_l$ represents an $l\times l$ identity matrix. However, there are some existing studies that have considered cross-correlation among ARIS elements. The cross-correlation results in the covariance matrix of $\mathbf{z}_{k,t}$ not being a diagonal matrix. In this study, we neglect the cross-correlation of $\mathbf{z}_{k,t}$ for the following two reasons. First, in the simulations in Sec. \ref{s5-2-1}, $\mathbf{z}_{k,t}$ undergoes path loss, resulting in a power at the RX that is far lower than $\mathbf{v}_k$. Therefore, neglecting cross-correlation does not exert a significant influence on the accuracy of the CRLB and simulation results. Second, simplifying the noise to additive white Gaussian noise can streamline the derivation and analysis of the CRLB. For these two reasons, the cross-correlation is neglected here.

Varying RIS phase configurations across symbol intervals produces distinct channel responses, enriching the information that can be captured about the ROI. Among all the multipath components from TX to RX, the TX-RIS-ROI-RX path, as shown by the red lines in Figure~\ref{f1-b}, carries essential imaging information, and can be extracted from $\mathbf{H}_k$ through channel estimation \cite{wang2025reconfigurable}. We refer to this path as the imaging path, whose corresponding channel measurements are employed for low-altitude imaging in Sec. \ref{s3}. 

\subsection{Channel Model}
\label{s2-2}

The imaging path starting from the $i$-th transmitting antenna during the $k$-th emission, scattered by the $t$-th RIS, and ultimately reaching the $j$-th receiving antenna can be expressed as \cite{huang2024ris}
\begin{equation}
h_{i,k,t,j}=g\mathbf{h}^{\text{T}}_{\text{TX}_{i},\text{s}_{t} } \mathbf{\Phi }_{k,t} \mathbf{H}_{\text{s}_{t},\text{v}}\text{diag}\left ( \mathbf{x}  \right )  \mathbf{h}_{\text{v},\text{RX}_{j} },
\label{eq2}
\end{equation}
where $g = \lambda / \sqrt{4\pi}$, and $\lambda$ is the wavelength of the center subcarrier. The channel between the $i$-th transmitting antenna and the $t$-th RIS is given as EQ.~\eqref{eq2-1} at the top of thia page, where $d_{\text{TX} _i,\text{s} _{t,m}} $ denotes the distance between the $i$-th transmitting antenna and the $m$-th element of the $t$-th RIS. $\mathbf{H}_{\text{s}_t,\text{v}}\in\mathbb{C}^{M_t\times N }$ and $\mathbf{h}_{\text{v},\text{RX}_j} \in\mathbb{C}^{N\times 1 }$ represent the channel between the $t$-th RIS and voxels, and between voxels and the $j$-th receiving antenna, respectively. They can be written in similar forms to EQ.~\eqref{eq2-1}.
\begin{figure*}[t]
\centering
\begin{equation}
\mathbf{h}_{\text{TX}_i,\text{s}_t}=\left [ \frac{1}{\sqrt{4\pi } d_{\text{TX} _i,\text{s} _{t,1}}}e^{-j2\pi \frac{d_{\text{TX} _i,\text{s} _{t,1}}}{\lambda } } ,\frac{1}{\sqrt{4\pi } d_{\text{TX} _i,\text{s} _{t,2}}}e^{-j2\pi \frac{d_{\text{TX} _i,\text{s} _{t,2}}}{\lambda } } ,\cdots,\frac{1}{\sqrt{4\pi } d_{\text{TX} _i,\text{s} _{t,M_t}}}e^{-j2\pi \frac{d_{\text{TX} _i,\text{s} _{t,M_t}}}{\lambda } } \right ]^{\text{T}}.
\label{eq2-1}
\end{equation}
\hrulefill
\end{figure*}

Since $h_{i,k,t,j}$ undergoes scattering through the ROI, it contains image information. Furthermore, RIS phase variation can result in sufficient measurements. According to EQ.~\eqref{eq2}, $h_{i,k,t,j}$ can be expressed as a linear function of $\mathbf{x}$, which is given as
\begin{equation}
h_{i,k,t,j}= \mathbf{a}^{\text{T}}_{i,k,t,j}\mathbf{x}, 
\label{eq12}
\end{equation}
where
\begin{equation}
\begin{aligned}
\mathbf{a}_{i,k,t,j}=\left [ g\mathbf{h}^{\text{T}}_{\text{TX}_{i},\text{s}_{t} }\mathbf{\Phi}_{k,t}\mathbf{H}_{\text{s}_{t},\text{v}}\text{diag}\left ( \mathbf{\mathbf{h}_{\text{v},\text{RX}_{j} }}  \right ) \right ]^{\text{T}}.
\end{aligned}
\label{eq13}
\end{equation}

\subsection{Power Consumption Model}
\label{s2-3}

The power consumption of the PRIS-aided system is given by \cite{zhou2023framework}
\begin{equation}
P_{\text{PRIS}}=P_{\text{TX,PRIS}}+\sum_{t=1}^{T}M_t P_{\text{c}},
\label{eq10}
\end{equation}
where $P_{\text{TX,PRIS}}$ is the transmit power of the TX in the PRIS-aided system. $P_{\text{c}}$ is the power consumed by the phase shifters and control circuits of the RIS elements, which is assumed to be identical for the $T$ RISs.

Compared with PRIS, ARIS additionally requires active transmit power, $P_{\text{M}_t }$, and the direct-current (DC) biasing power, $P_{\text{DC}}$, which is used to drive amplifiers. Consequently, the power consumption of the ARIS-aided system is given by \cite{zhou2023framework}
\begin{equation}
P_{\text{ARIS}}=P_{\text{TX,ARIS}}+\sum_{t=1}^T \left [ P_{\text{M}_t}+ M_t \left( P_{\text{c}}+P_{\text{DC}}\right) \right ] .
\label{eq9}
\end{equation}
where $P_{\text{TX,ARIS}}$ is the transmit power of the TX in the ARIS-aided system. $P_{\text{M}_t}$ denotes the active transmit power of the $t$-th ARIS, which can be expressed as \cite{zhang2022active}
\begin{equation}
\begin{aligned}
P_{\text{M}_t}=\mathbb{E}\left \{ \left \|   \left ( g\sqrt{P_{\text{TX}}}\mathbf{s}_k^{\text{T}}\mathbf{H}_{\text{TX},\text{s}_{t} }+\mathbf{z}_{k,t} \right ) \mathbf{\Phi} _{k,t} \right \| _2^2 \right \} \\
=\left \|   g\sqrt{P_{\text{TX,ARIS}}}\mathbf{s}_k^{\text{T}}\mathbf{H}_{\text{TX},\text{s}_{t} }  \mathbf{\Phi} _{k,t}\right \| _2^2+\sum_{m=1}^{M_t} \left | \omega _{k,t,m} \right |^2 \sigma ^2_{\text{RIS} } ,
\end{aligned}
\label{eq7}
\end{equation} 
where $\mathbf{H}_{\text{TX},\text{s}_{t} } =\left [ \mathbf{h}_{\text{TX}_1,\text{s}_t },\cdots,  \mathbf{h}_{\text{TX}_{N_{\text{TX} }},\text{s}_t }\right ] ^{\text{T} }\in\mathbb{C}^{N_{\text{TX}}\times M_t}  $ represents the channel between the TX and the $t$-th RIS. 

\section{Problem Formulation and Imaging Algorithms}
\label{s3}

\subsection{Problem Formulation}
\label{s3-1}

By transmitting and receiving signals, varying the RIS phase configurations, and conducting channel estimation, we can acquire the CSI measurement of the imaging path, which can be expressed as
\begin{equation}
y_{i,k,t,j}= h_{i,k,t,j}+n_{i,k,t,j} ,
\label{eq11}
\end{equation}
where $n_{i,k,t,j}$ is the additive noise originating from channel estimation, multi-path effects, and scattering more than three bounces.

By stacking the measurements of all transmitting antennas, RISs, receiving antennas, and symbol intervals, we can derive 
\begin{equation}
\mathbf{y} =\mathbf{A} \mathbf{x} + \mathbf{n},
\label{eq14}
\end{equation}
where
\begin{equation}
\begin{aligned}
\mathbf{y}=\left [ y_{1,1,1,1},\cdots ,y_{N_{\text{TX} },K,T,N_{\text{RX}}} \right ]^{\text{T}} \in\mathbb{C} ^{L\times 1},\\
\mathbf{A}=\left [ \mathbf{a}_{1,1,1,1},\cdots,\mathbf{a}_{N_{\text{TX} },K,T,N_{\text{RX}}} \right ]^{\text{T}} \in\mathbb{C} ^{L\times N},\\
\mathbf{n}=\left [ n_{1,1,1,1},\cdots,n_{N_{\text{TX} },K,T,N_{\text{RX}}} \right ]^{\text{T}} \in\mathbb{C} ^{L\times 1}.
\end{aligned}
\label{eq15}
\end{equation}
Here, we have $L=KTN_{\text{TX}}N_{\text{RX}}$, $\mathbf{y}$ denotes the CSI measurements, and $\mathbf{A}$ represents the sensing matrix obtained from the channel model and RIS phase configurations. Given that $\mathbf{x}$ exhibits high sparsity, we can reconstruct it from $\mathbf{y}$ using $\mathbf{A}$ and formulate the imaging problem as \cite{learned_off_grid}
\begin{equation}
\hat{\mathbf{x}} =\underset{\mathbf{x}}{{\text{argmin}} } \left \| \mathbf{x} \right \| _1,\quad\text{s.t.}\left \| \mathbf{y}-\mathbf{A}\mathbf{x}    \right \| _2^2\le  \varepsilon ,
\label{p1}
\end{equation}
where $\varepsilon $ is a small threshold ensuring reconstruction accuracy.

\subsection{Imaging Algorithms}
\label{s3-2}
The reconstruction problem can be efficiently solved using CS theory. Several algorithms have been proposed to tackle this problem, providing multiple options. In this study, we employ the SP algorithm, which is a classical algorithm that first estimates the support of $\mathbf{x}$, and then, estimates the corresponding values. In the SP algorithm, the residual signal computation is defined as \cite{dai2009subspace}
\begin{equation}
\mathbf{y}_{\text{res}}\left ( \mathbf{u} \right ) =\mathbf{y}-\mathbf{A}_\mathbf{u} f_{\text{LS}}\left ( \mathbf{y},\mathbf{A}_\mathbf{u} \right ) ,
\label{eq15-1}
\end{equation}
where $\mathbf{u} = [u_1, \dots, u_S]^{\text{T}}$ represents the support set of $\mathbf{x}$. Since the exact number of UAVs in the aerial space is unknown, the algorithm uses a prior-based sparsity value $S$, which denotes the number of non-zero elements in $\mathbf{x}$. In real deployments, we can set $S$ to a large value that meets the sparsity requirement of the SP algorithm, when the ground-truth sparsity is unknown a priori. The value of $S$ can be chosen with the aid of prior information and experience. When $S$ input to the algorithm is larger than the ground-truth number of UAVs, the SP algorithm may still achieve near-perfect estimation results. This setting ensures that voxels containing targets are detected, while other voxels are estimated as zero. The non-zero values in $\mathbf{x}$ can be expressed as $\mathbf{x}_{\mathbf{u} } = [x_{u_1}, \dots, x_{u_S}]^{\text{T}}$. Decomposing the sensing matrix $\mathbf{A}$ into column vectors as $\mathbf{A}=\left [ \mathbf{p}_1,\cdots ,  \mathbf{p}_N\right ] $. $\mathbf{A}_\mathbf{u}=\left [ \mathbf{p}_{u_1},\cdots ,  \mathbf{p}_{u_S}\right ]$ is the submatrix of $\mathbf{A}$, corresponding to $\mathbf{x}_{\mathbf{u} }$. The function $f_{\text{LS}}\left ( \mathbf{y},\mathbf{A}_\mathbf{u} \right )$ computes the least squares (LS) estimate of $\mathbf{x}$ using $\mathbf{y}$ and $\mathbf{A}_{\mathbf{u}}$. The function $f_{\text{sel,S}}\left ( \mathbf{y},\mathbf{A} \right ) $ selects the indices corresponding to the largest $S$ absolute value of $\mathbf{A}^{\text{H}} \mathbf{y}$, effectively selecting the potential support. 

The SP algorithm starts by generating an initial support $\mathbf{u}_0$ using $f_{\text{sel,S}}\left ( \mathbf{y},\mathbf{A} \right ) $ and computing the corresponding residual $\mathbf{y}_{\text{res}}\left ( \mathbf{u}_0 \right )$. In the $c$-th iteration, the following steps are conducted:
\begin{enumerate}
	\item Augment $\mathbf{u}_{c-1}$ by adding the indices obtained from $f_{\text{sel,S}}\left ( \mathbf{y}_{\text{res}}\left ( \mathbf{u}_{c-1} \right ),\mathbf{A} \right ) $, forming $\hat{\mathbf{u}}_c$.
	\item Refine the support as $\mathbf{u}_c=f_{\text{sel,S}}\left ( \mathbf{y},\mathbf{A}_{\hat{\mathbf{u}}_c} \right )$.
	\item Compute the new residual signal $\mathbf{y}_{\text{res}}\left ( \mathbf{u}_{c} \right )$.
\end{enumerate}
The process repeats until the residual signal energy falls below the threshold $\varepsilon$ or $\mathbf{u}$ stabilizes. Finally, the estimated non-zero values in $\mathbf{x}$ are computed as $f_{\text{LS}}\left ( \mathbf{y},\mathbf{A}_{\mathbf{u}_{\text{final}}} \right )$, where $\mathbf{u}_{\text{final}}$ denotes the final estimated support set.

According to \cite{dai2009subspace}, the total complexity of the SP algorithm is given by $\mathcal{O}\left ( L(N+S^2)S \right ) $, and it can be further reduced to $\mathcal{O}\left ( LN\log_{}{S}  \right ) $ for very sparse signals. Compared with another commonly used greedy algorithm, the Orthogonal Matching Pursuit (OMP) algorithm, whose complexity is $\mathcal{O}(LNS)$ \cite{10621011}, the SP algorithm has a comparable complexity and even a lower one when $\mathbf{x}$ is extremely sparse. Since both the SP and OMP algorithms use the LS method to solve for $\mathbf{x}$ after identifying the support set, they achieve similar accuracy under ideal imaging conditions. When running the imaging process on a Central Processing Unit (CPU) of the model 12th Gen Intel(R) Core(TM) i7-1255U, the average runtime of the SP algorithm and the OMP algorithm is 0.075 seconds and 0.127 seconds, respectively. Both algorithms can complete the solution quickly, with the SP algorithm being more optimal.

\section{CRLB for Imaging}
\label{s4}
In this section, we derive the CRLB of $\mathbf{x}$, then analyze its expectation when employing random RIS phase configurations, and finally discuss the influence of various system parameters.

\subsection{CRLB Derivation}
\label{s4-1}

Assuming that the transmit power $P_{\text{TX}}$ is equally allocated to the transmitting antennas, the CRLB for unbiased estimation of the $r$-th element in $\mathbf{x}_{\mathbf{u} }$ is given by \cite{li2024radio}
\begin{equation}
\text{CRLB}\left (  x_{u_r} \right ) = \left ( \gamma \mathbf{A}_\mathbf{u} ^{\text{H} }\mathbf{A}_\mathbf{u}  \right ) ^{-1}_{r,r},1\le r\le S ,
\label{eq16}
\end{equation}
where $\gamma$ denotes the reciprocal of the noise power. The average CRLB of all the $S$ elements is given by
\begin{equation}
\text{CRLB}=\frac{1}{S} \text{Tr} \left \{ \left ( \gamma \mathbf{A}_\mathbf{u}^{\text{H} }\mathbf{A}_\mathbf{u} \right ) ^{-1} \right \},
\label{eq31}
\end{equation}
where $\text{Tr} $ denotes the trace operator.
Considering that the locations of low-altitude targets are random, we derive the CRLB for each voxel. When $S=1$ and $u_1=n$, the CRLB of the scattering coefficient $x_n$ is given by
\begin{equation}
C_n=\left ( \gamma \mathbf{p}_n^{\text{H} }\mathbf{p}_n  \right )^{-1}.
\label{eq17}
\end{equation}

In ARIS-aided systems, we assume that all ARIS units operate at the maximum amplification factor $\alpha_{\max}$, i.e., $\left | \omega_{k,t,m} \right | ^2=\alpha_{\max}$. For convenience, we define $a \triangleq \left| \omega_{k,t,m} \right|^2$, such that $a=1$ for PRIS-aided systems, and $a=\alpha_{\max}$ for ARIS-aided systems. Under this configuration, the elements of $\mathbf{A}$ can be uniformly expressed for both ARIS and PRIS-aided systems. $\mathbf{A}$ consists of $L$ rows, where each row corresponds to the index set $\left ( i,k,t,j \right )$. $\mathbf{A}_{\left ( i,k,t,j \right ),n }$ represents the $n$-th element of the row corresponding to $\left ( i,k,t,j \right )$, given as
\begin{equation}
\begin{aligned}
\mathbf{A}_{\left ( i,k,t,j \right ),n }=&g\frac{e^{-j2\pi \frac{d_{\text{v}_n,\text{RX} _j}}{\lambda } }}{\sqrt{4\pi } d_{\text{v}_n,\text{RX} _j}}\times\\
&\sum_{m=1}^{M_t} \omega _{k,t,m}\frac{e^{-j2\pi\frac{d_{\text{TX}_i,\text{s}_{t,m}}+d_{\text{s}_{t,m},\text{v}_n}}{\lambda} }}{4\pi d_{\text{TX}_i,\text{s}_{t,m}}d_{\text{s}_{t,m},\text{v}_n}} ,
\end{aligned}
\label{eq19}
\end{equation}
where $d_{\text{v}_n,\text{RX} _j}$, $d_{\text{TX}_i,\text{s}_{t,m}}$, and $d_{\text{s}_{t,m},\text{v}_n}$ represent the distance between the $n$-th voxel and the $j$-th receiving antenna, between the $i$-th transmitting antenna and the $m$-th element of the $t$-th RIS, and between the $m$-th element of the $t$-th RIS and the $n$-th voxel, respectively. Thus, the CRLB of $x_n$ is given as EQ.~\eqref{eq20} at the top of the next page.
\begin{figure*}[t]
\centering
\begin{equation}
\begin{aligned}
C_n=\left ( \sum_{i,k,t,j}^{} \frac{\gamma  g^2  }{4\pi d_{\text{v}_n,\text{RX} _j}^2}\times\left \|\sum_{m=1}^{M_t} \omega _{k,t,m}\frac{e^{-j2\pi\frac{d_{\text{TX}_i,s_{t,m}}+d_{s_{t,m},v_n}}{\lambda} }}{4\pi d_{\text{TX}_i,s_{t,m}}d_{s_{t,m},\text{v}_n}}   \right \|_2^2 \right )^{-1} .
\end{aligned}
\label{eq20}
\end{equation}
\end{figure*}

Assuming that the phase shifts of all RIS elements are mutually independent and uniformly distributed, we have $\mathbb{E} \left \{ \omega _{k_1,t_1,m_1}^{* }\omega _{k_2,t_2,m_2} \right \} =0$, where $\left ( k_1,t_1,m_1 \right ) \ne \left ( k_2,t_2,m_2 \right )$. Consequently, the expectation of $C_n$ is expressed as \eqref{eq21} at the top of the next page. 
\begin{figure*}[t]
\centering
\begin{equation}
\begin{aligned}
\mathbb{E} \left \{ C_n \right \}&=\left ( \sum_{i,k,t,j}^{} \frac{\gamma  g^2  }{4\pi d_{\text{v}_n,\text{RX} _j}^2}\sum_{m=1}^{M_t} \frac{a}{16\pi^2 d_{\text{TX}_i,\text{s}_{t,m}}^2 d_{\text{s}_{t,m},\text{v}_n}^2}  \right ) ^{-1}=\frac{64\pi^3}{a\gamma g^2} \left ( \sum_{i,k,t,j}^{} \frac{1}{d_{\text{v}_n,\text{RX} _j}^2} \sum_{m=1}^{M_t} \frac{1}{d_{\text{TX}_i,\text{s}_{t,m}}^2 d_{\text{s}_{t,m},\text{v}_n}^2} \right ) ^{-1}\\
&=\frac{64\pi^3}{a\gamma g^2 K }\left [ \left ( \sum_{j=1}^{N_{\text{RX} } }\frac{1}{d_{\text{v}_n,\text{RX} _j}^2}  \right )\left ( \sum_{i=1}^{N_{\text{TX} }}\sum_{t=1}^{T} \sum_{m=1}^{M_t} \frac{1}{d_{\text{TX}_i,\text{s}_{t,m}}^2 d_{\text{s}_{t,m},\text{v}_n}^2}   \right ) \right ]  ^{-1}.
\label{eq21}
\end{aligned}
\end{equation}
\hrulefill
\end{figure*}
To analyze the influence of system parameters on $\mathbb{E}\left \{ C_n \right \}$, we decompose it into the product of three distinct factors as follows
\begin{equation}
\mathbb{E}\left \{ C_n \right \}=\eta\kappa \mu,
\label{eq22}
\end{equation}
where 
\begin{equation}
\eta =\frac{64\pi^3}{a\gamma g^2 K },
\label{eq23}
\end{equation}
\begin{equation}
\kappa =\left ( \sum_{j=1}^{N_{\text{RX} } }\frac{1}{d_{\text{v}_n,\text{RX} _j}^2}  \right )^{-1},
\label{eq24}
\end{equation}
\begin{equation}
\mu =\left ( \sum_{i=1}^{N_{\text{TX} }}\sum_{t=1}^{T} \sum_{m=1}^{M_t} \frac{1}{d_{\text{TX}_i,\text{s}_{t,m}}^2 d_{\text{s}_{t,m},\text{v}_n}^2}   \right ) ^{-1}.
\label{eq25}
\end{equation}
Here, $\eta$ is inversely proportional to the number of sensing symbol intervals, $K$, and the amplification factor of the RIS, $a$. Alternatively, $\kappa$ is related to the channel between the ROI and the RX, and $\mu$ is related to the TX-RIS-ROI path.

However, EQ.~\eqref{eq24} and EQ.~\eqref{eq25} involve highly complex distance calculations. Thus, we find an approximation of $\mathbb{E} \left \{ C_n \right \}$ to simplify the expression. Considering that $d_{s_{t,m},v_n}$ is much larger than the spacing of RIS elements, we approximate it by the distance between the $n$-th voxel and the center of the $t$-th RIS, denoted as $d_{s_t,v_n}$. Similarly, we replace $d_{\text{TX}_i,s_{t,m}}$ and $d_{v_n,\text{RX} _j}$ with the distance between the center of TX and the center of the $t$-th RIS, $d_{\text{TX},s_t}$, and the distance between the $n$-th voxel and the center of RX, $d_{v_n,\text{RX}}$, respectively. Consequently, $\mathbb{E} \left \{ C_n \right \}$ can be approximated as
\begin{equation}
\mathbb{E} \{ \tilde{C}_{n}  \} =\eta\tilde{\kappa}\tilde{  \mu   } ,
\label{eq26}
\end{equation}
where
\begin{equation}
\tilde{\kappa}=\frac{d_{\text{v}_n,\text{RX}}^2}{N_{\text{RX}}},
\label{eq27}
\end{equation}
\begin{equation}
\tilde{\mu} =\left ( \sum_{t=1}^{T} \frac{M_{t}N_{\text{TX}}}{d_{\text{TX},\text{s}_t}^2 d_{\text{s}_t,\text{v}_n}^2}   \right ) ^{-1}.
\label{eq28}
\end{equation}
According to EQ.~\eqref{eq27}, $\mathbb{E} \{ \tilde{C}_{n} \}$ is proportional to the square of the distance between the voxel and the RX, i.e., $d_{v_n,\text{RX}}^2$. According to EQ.~\eqref{eq28}, $\mathbb{E} \{ \tilde{C}_{n}  \}$ is positively correlated with $d_{\text{TX},s_t}^2 d_{\text{s}_t,\text{v}_n}^2$.

\begin{figure}[!ht]
\centering
\includegraphics[width=0.47\textwidth]{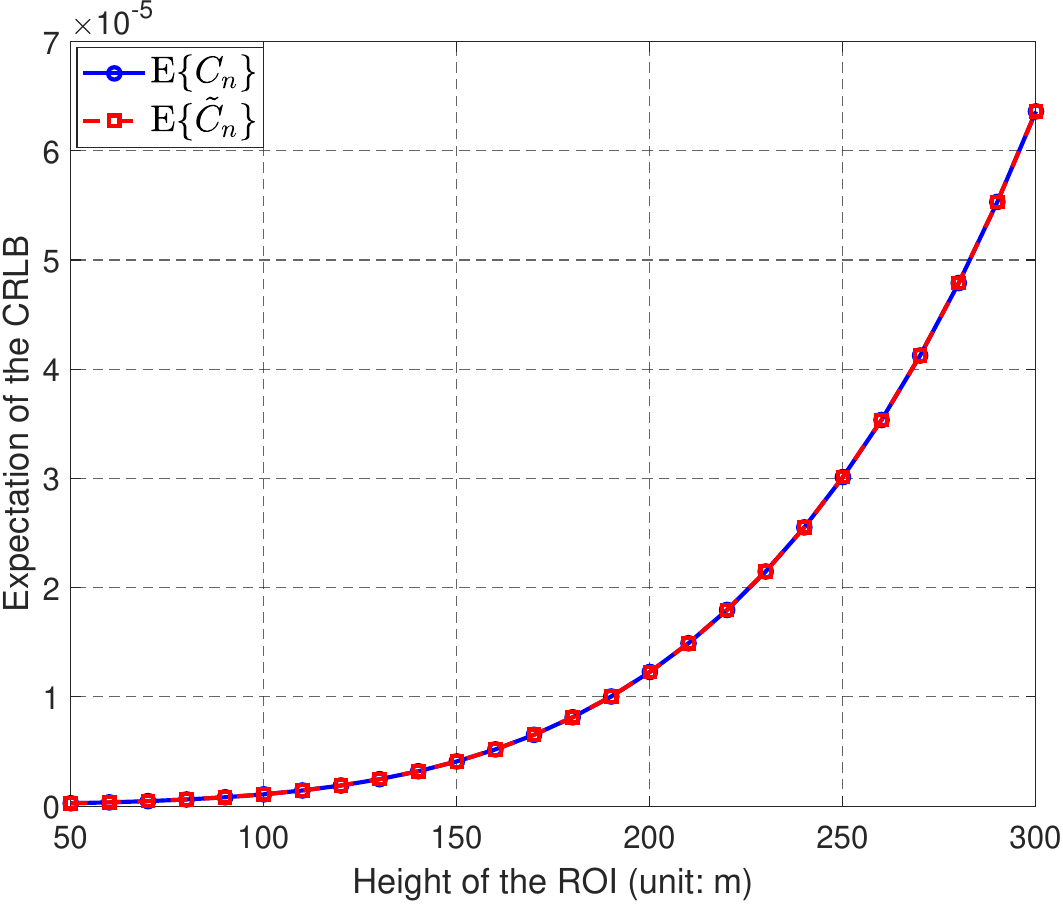}
\caption{Comparison between $\mathbb{E} \{ C_n  \}$ and $\mathbb{E} \{ \tilde{C}_n \}$.}
\label{fig3}
\end{figure}

Under the simulation settings in Sec.~\ref{s5}, Figure~\ref{fig3} compares the values of $\mathbb{E} \{ C_n \}$ and $\mathbb{E} \{ \tilde{C}_n  \}$. The closely matched lines show that $\mathbb{E} \{ \tilde{C}_n  \}$ is a tight approximation of $\mathbb{E} \{ C_n  \}$, and the properties of $\mathbb{E} \{ C_n  \}$ can be properly represented by $\mathbb{E} \{ \tilde{C}_n  \}$.

\subsection{CRLB analysis}
Having derived the expression of $\mathbb{E} \{ \tilde{C}_n  \}$ in Sec.~\ref{s4-1}, this subsection analyzes how various system parameters influence it, based on the simulation parameters in Sec.~\ref{s5}.
\label{s4-2}
\subsubsection{Target Position}
\label{s4-2-1}

\begin{figure}[t]
\centering
\includegraphics[width=0.47\textwidth]{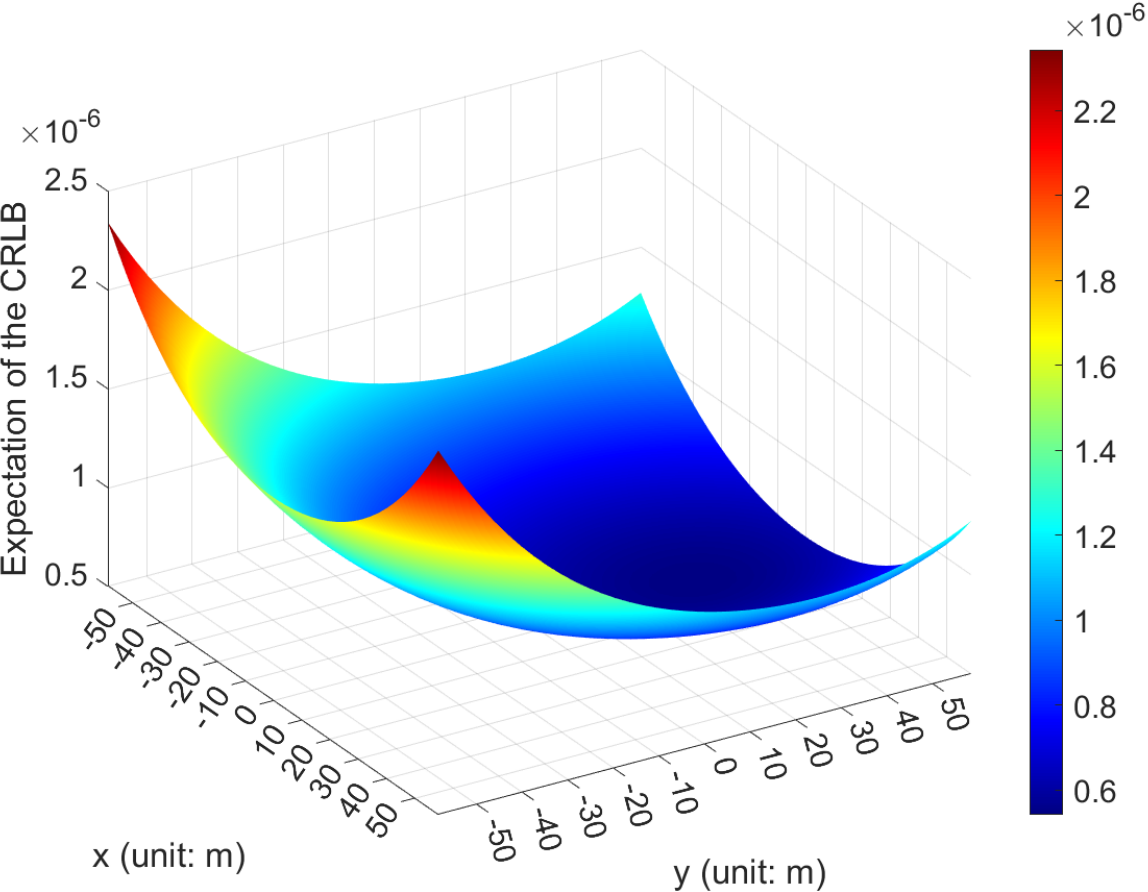}
\caption{$\mathbb{E} \{\tilde{C}_n  \}$ versus voxel position when the RX is centered at $\left[0,60\text{m},30\text{m}\right]^{\text{T}}$.}
\label{fig5}
\end{figure}

To investigate the detection performance of aerial targets located at different positions, we present the expectation of the CRLB, $\mathbb{E} \{\tilde{C}_n  \}$, of all voxels in Figure~\ref{fig5}. The simulation results show that the ROI interior exhibits low $\mathbb{E} \{\tilde{C}_n  \}$, which corresponds to high imaging performance. Alternatively, the exterior region obtains high $\mathbb{E} \{\tilde{C}_n  \}$, meaning that the sensing performance for targets far from the ROI center is degraded. Additionally, $\mathbb{E} \{\tilde{C}_n  \}$ surface is symmetric along the x-axis, whereas this symmetry does not hold along the y-axis. $\mathbb{E} \{\tilde{C}_n  \}$ in the negative y-axis is higher than that in the positive y-axis. This asymmetry in $\mathbb{E} \{\tilde{C}_n  \}$ is due to the influence of $d_{\text{v}_n,\text{RX}}^2$ in EQ.~\eqref{eq27}, which indicates that as the distance between the voxel and the RX increases, $\mathbb{E} \{\tilde{C}_n  \}$ also increases, leading to a decline in estimation accuracy.

\begin{figure}[t]
\centering
\includegraphics[width=0.47\textwidth]{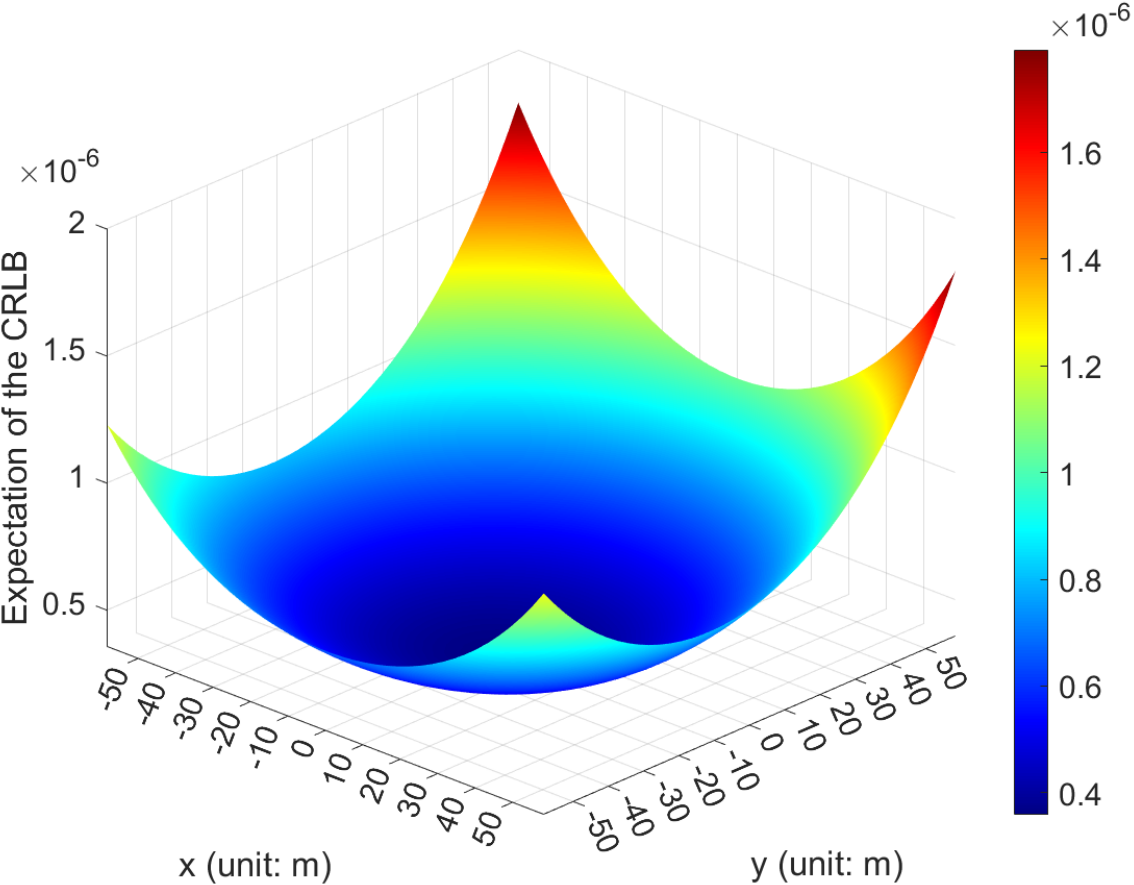}
\caption{$\mathbb{E} \{\tilde{C}_n  \}$ versus voxel position when the RX is centered at $\left[0,0,30\,\text{m}\right]^{\text{T}}$.}
\label{fig6}
\end{figure}

We reposition the RX to $\left[0,0,30\text{m}\right]^{\text{T}}$ to achieve symmetric voxels equidistant to the RX, thereby reducing the impact of $d_{\text{v}_n,\text{RX}}^2$. The simulation results are shown in Figure~\ref{fig6}. While $\mathbb{E} \{\tilde{C}_n  \}$ remains asymmetric along the y-axis, $\mathbb{E} \{\tilde{C}_n  \}$ in the positive y-axis is higher than that in the negative y-axis, which is different from Figure~\ref{fig5}. This difference in $\mathbb{E} \{\tilde{C}_n  \}$ arises from $\tilde{\mu} = \left( \sum_{t=1}^{T} \frac{M_{t}N_{\text{TX}}}{d_{\text{TX},\text{s}_t}^2 d_{\text{s}_t,\text{v}_n}^2} \right)^{-1}$. Since four RISs are deployed symmetrically about the ROI, the term $d_{\text{s}_t,\text{v}_n}^2$ is symmetric. However, the TX is deployed at $\left[0,-60\text{m},30\text{m}\right]^{\text{T}}$, which means that the term $d_{\text{TX},\text{s}_t}^2$ is asymmetric, destroying the symmetry of $\tilde{\mu}$. The term $d_{\text{TX},\text{s}_t}^2$ results in that the four RISs impose different influences on $\mathbb{E} \{\tilde{C}_n  \}$, where the two RISs closer to the TX have greater impacts on it. Consequently, $\mathbb{E} \{\tilde{C}_n  \}$ is positively correlated with the distances between the voxel and four RISs, but is more strongly influenced by the RISs closer to the TX. In Figure~\ref{fig6}, voxels in the positive y-axis having higher $\mathbb{E} \{\tilde{C}_n  \}$. It is due to the fact that they are farther from these two RISs.

By comparing Figure~\ref{fig5} and Figure~\ref{fig6}, it can be concluded that the distance between the voxel and the RX is the primary factor that influences $\mathbb{E} \{\tilde{C}_n  \}$. Specifically, when the distance between the voxel and the RX increases, $\mathbb{E} \{\tilde{C}_n  \}$ increases. Additionally, the distances between the voxels and RISs which are closer to the TX also affect $\mathbb{E} \{\tilde{C}_n  \}$. When the distances between the voxels and these RISs increase, $\mathbb{E} \{\tilde{C}_n  \}$ grows. However, the influence of the distance between the voxel and the RX is more significant than the influence of the distances between the voxels and RISs.

Since $\mathbb{E} \{\tilde{C}_n  \}$ varies with different target positions, we define the average of $\mathbb{E} \{\tilde{C}_n  \}$ over all the $N$ voxels as
\begin{equation}
\mathbb{E}  \{ \bar{\tilde{C} }  \} =\frac{1}{N} \sum_{n=1}^{N} \mathbb{E}  \{ \tilde{C}_n  \},
\label{eq28-1}
\end{equation}
and we examine the influence of different system parameters on it in the remaining part of this subsection.

\subsubsection{RX position}
\label{s4-2-2}

\begin{figure}[t]
\centering
\includegraphics[width=0.47\textwidth]{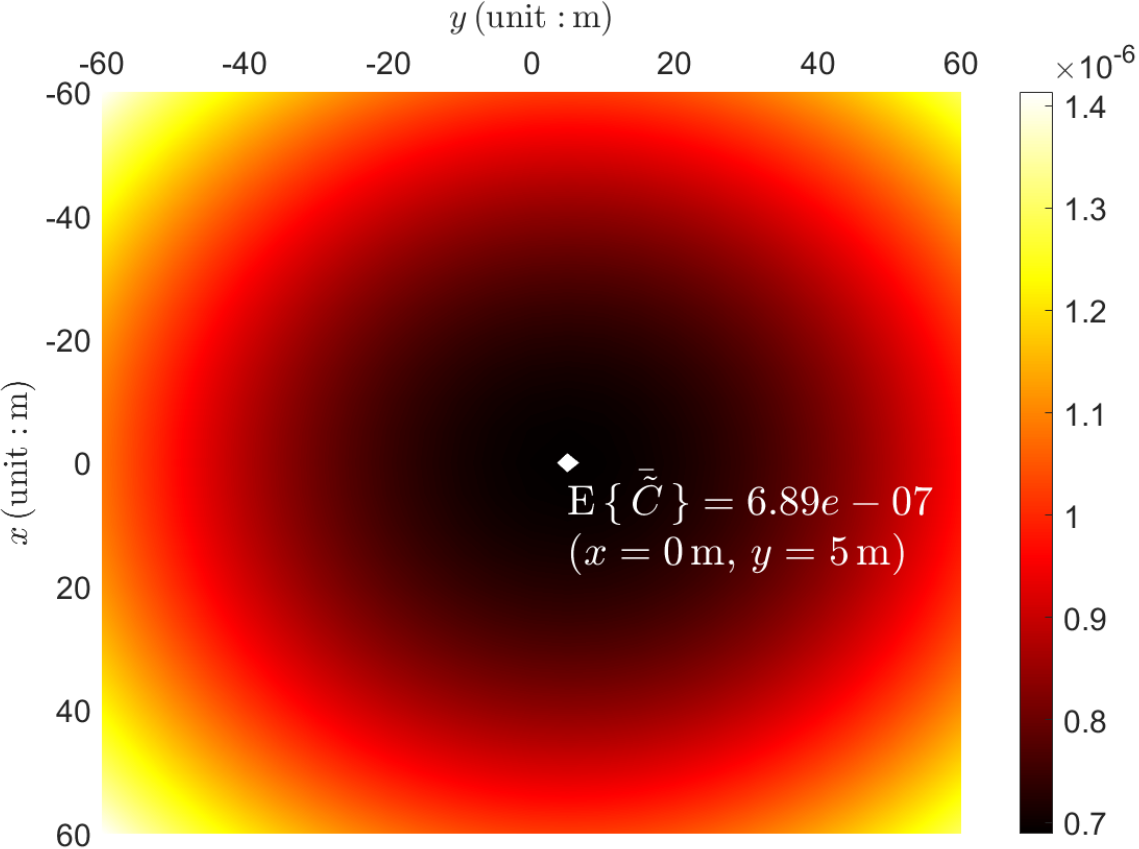}
\caption{$\mathbb{E}  \{ \bar{\tilde{C} }   \} $ versus the RX position.}
\label{fig7}
\end{figure}

According to Sec.~\ref{s4-2-1}, the distances from the voxels to the RX exert significant influences on $\mathbb{E} \{\tilde{C}_n  \}$. Consequently, to minimize $\mathbb{E}  \{ \bar{\tilde{C} }   \} $, the RX location should be properly selected. Figure~\ref{fig7} illustrates $\mathbb{E}  \{ \bar{\tilde{C} }   \}$, where the RX position varies across the $120\text{m} \times 120\text{m}$ horizontal grid while maintaining a constant height of $30\text{m}$. We note that the x- and y-axes in Figure~\ref{fig5} and Figure~\ref{fig6} represent the coordinates of the voxels where the targets are located, whereas the x- and y-axes in Figure~\ref{fig7} represent the coordinates of the RX.

$\mathbb{E}  \{ \bar{\tilde{C} }   \}$ reaches the minimum value when the RX is positioned at $\left[0,5\text{m},30\text{m}\right]^{\text{T}}$. Theoretically, the sum of the squared distances from all voxels to the RX at $\left[0,0,30\,\text{m}\right]^{\text{T}}$ should be minimized. However, owing to the influence of $\tilde{\mu}$, $\mathbb{E} \{ \bar{\tilde{C} }   \}$ exhibits larger values for voxels in the positive y-axis. Consequently, the optimal RX location is marginally shifted toward the positive y-axis, which is located at $\left[0,5\text{m},30\text{m}\right]^{\text{T}}$ according to our simulations. It indicates that the optimal horizontal position of the RX lies near the center of the ROI, under the scenarios considered in this study.

\subsubsection{TX position}
\label{s4-2-3}

\begin{figure}[t]
\centering
\includegraphics[width=0.47\textwidth]{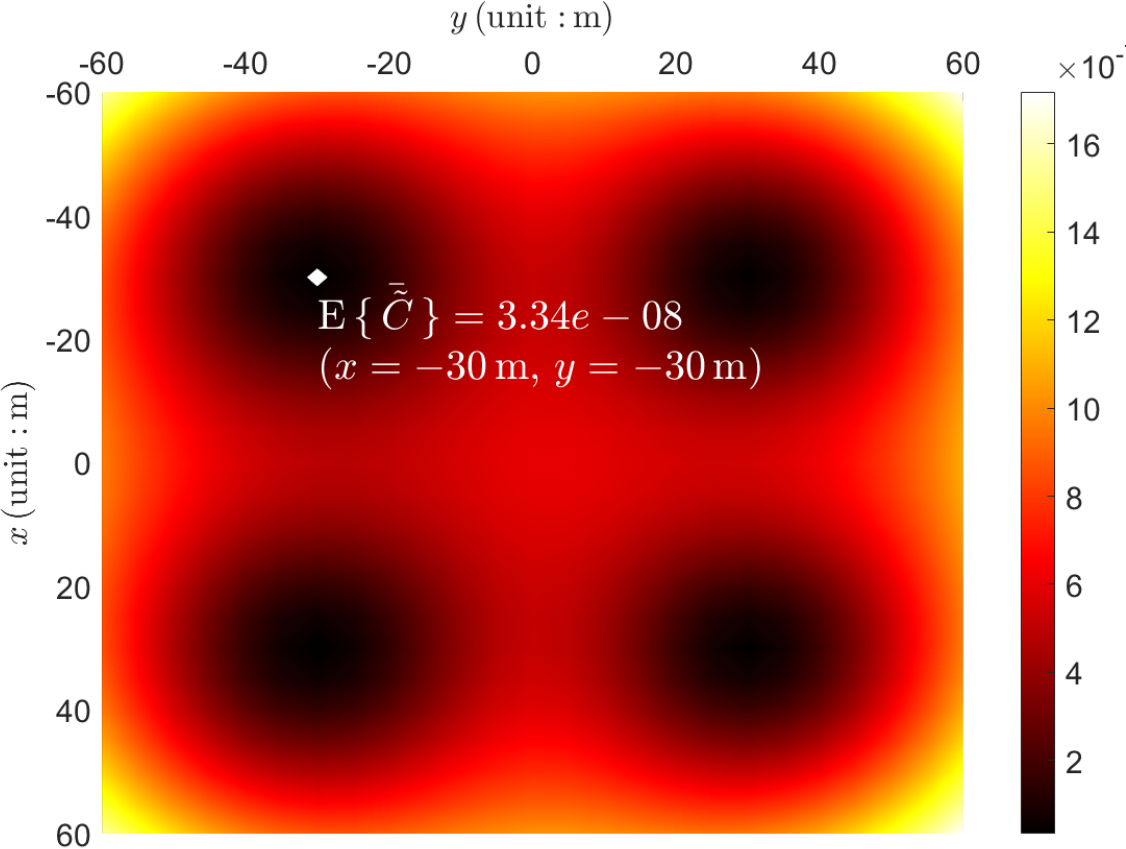}
\caption{$\mathbb{E}  \{ \bar{\tilde{C} }   \} $ versus the TX position.}
\label{fig8}
\end{figure}

The position of the TX influences the distance from the TX to the RISs. Figure~\ref{fig8} illustrates $\mathbb{E}  \{ \bar{\tilde{C} }   \} $, where the TX position varies across the $120\text{m} \times 120\text{m}$ horizontal grid while maintaining a constant height of $30\text{m}$. $\mathbb{E} \{\tilde{C}_n  \}$ reaches the local minimum value when the TX is located at $\left [ \pm 30\text{m},\pm 30\text{m},30\text{m} \right ] ^{\text{T}}$, directly above one of the centers of the RISs, since $\tilde{\mu}$ is easily dominated by the smallest $d_{\text{TX},\text{s}_t}^2 d_{\text{s}_t,\text{v}_n}^2$. Specifically, when the TX is exactly above the $t_0$-th RIS, the value of $d_{\text{TX},\text{s}_{t_0}}^2 d_{\text{s}_{t_0},\text{v}_n}^2$ becomes significantly lower than that of the other three RISs. Thus, $\mathbb{E}  \{ \bar{\tilde{C} }   \} $ is dominated by $d_{\text{TX},\text{s}_{t_0}}^2 d_{\text{s}_{t_0},\text{v}_n}^2$. Consequently, $\mathbb{E}  \{ \bar{\tilde{C} }   \} $ reaches its local minimum value when the TX is located exactly above one of the RISs. It indicates that placing the TX close to one of the RISs yields favorable imaging performance.

\subsubsection{RIS positions}
\label{s4-2-4}

\begin{figure}[t]
\centering
\includegraphics[width=0.47\textwidth]{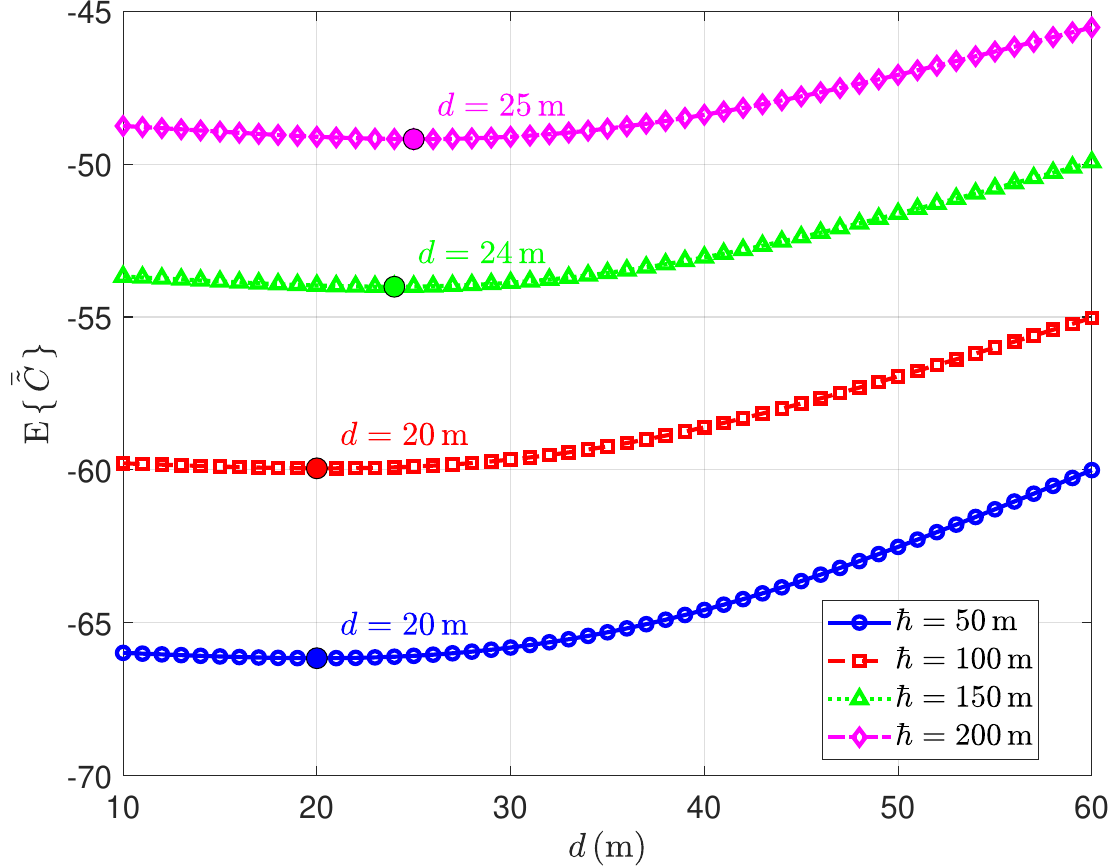}
\caption{$\mathbb{E}  \{ \bar{\tilde{C} }   \} $ versus $d$.}
\label{fig9}
\end{figure}

Given the complexity of optimizing the positions of multiple RISs \cite{encinas2025riloco}, we employ four RISs symmetrically positioned at $\left[\pm d, \pm d, 25\,\text{m}\right]^{\text{T}}$, simplifying the following analysis. Figure~\ref{fig9} illustrates $\mathbb{E} \{ \bar{\tilde{C} }   \}$ for different $d$ and ROI heights, $\hbar$.

The results indicate that as $d$ increases, $\mathbb{E} \{ \bar{\tilde{C} } \}$ initially exhibits a gradual decline trend, followed by a fast increase. The optimal $d$ corresponding to the minimum $\mathbb{E}  \{ \bar{\tilde{C} }   \} $ has been marked in the figure, which varies from $20\text{m}$ to $25\text{m}$ for different ROI heights. Notably, as the ROI height $h$ increases, the optimal $d$ shifts slightly. When $d$ is below the optimal value, variations in $\mathbb{E} \{ \bar{\tilde{C} }   \}$ are minimal. To explain these results, conducting theoretical analysis using EQ.~\eqref{eq28} is challenging, as altering the RIS positions simultaneously affects the lengths of TX-RIS and RIS-ROI paths. Nevertheless, the results in Figure~\ref{fig9} empirically demonstrate the influence of path lengths on imaging accuracy. A comprehensive analysis integrating additional metrics and factors, such as aperture size, will be presented in the subsequent section.

\section{Simulation Results}
\label{s5}
In this section, we conduct simulations using the SP algorithm and analyze the sensing performance of the proposed systems. 

\subsection{Experiment Setup and Evaluation Metrics}
\label{s5-1}
\begin{figure*}[t]
\centering
\includegraphics[width=0.9\textwidth]{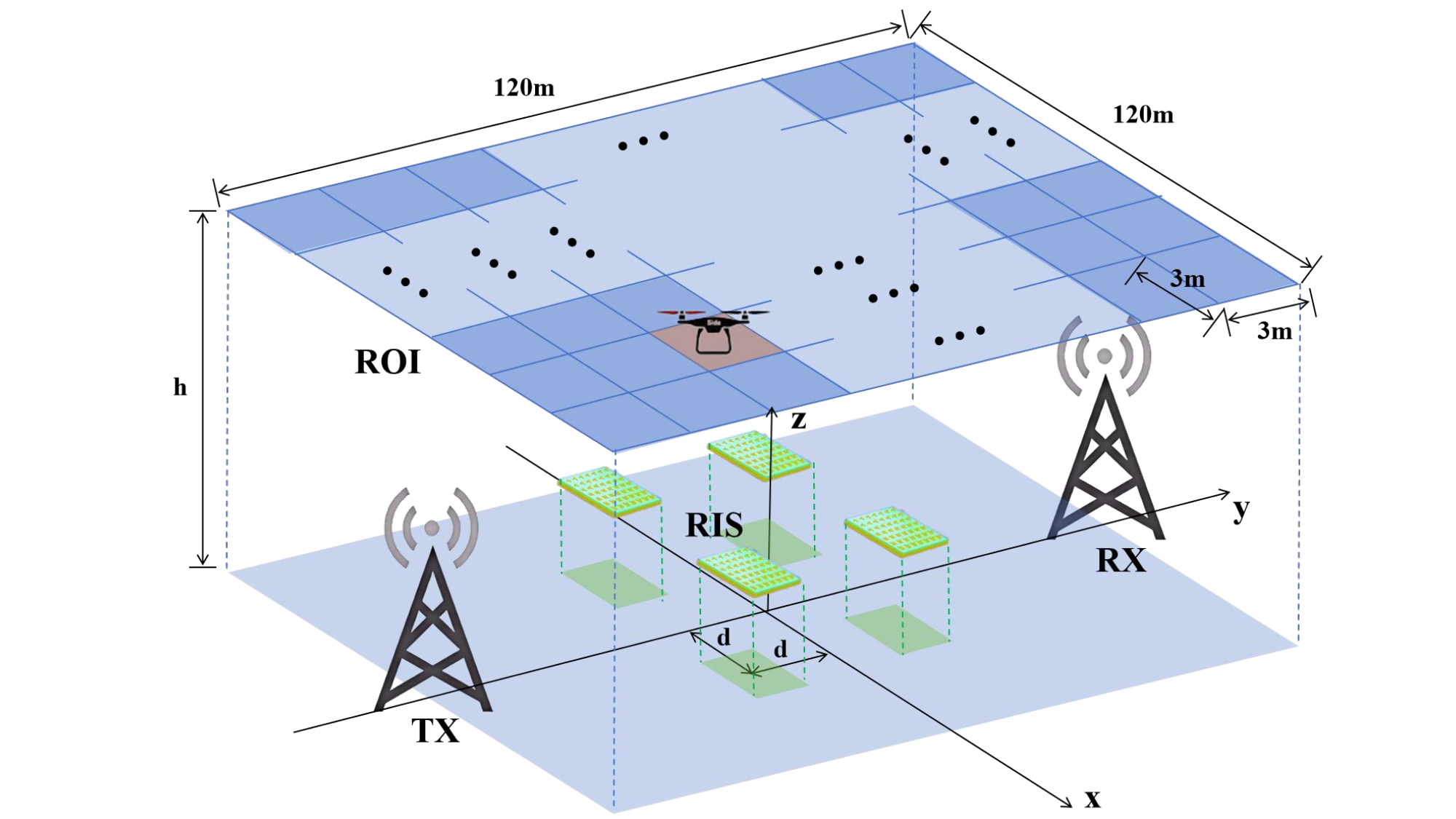}
\caption{Illustration of scenario in the simulations.}
\label{fig4}
\end{figure*}

We consider the simulation scenario shown in Figure~\ref{fig4}. The ROI is centered at $\left[0,0,\hbar\right]^{\text{T}}$, where $50\text{ m} \le \hbar \le 400\text{ m}$. The dimensions of the ROI are $120\text{ m} \times 120\text{ m} \times 3\text{ m}$. It is divided into $40 \times 40 \times 1$ voxels, each of size $3\text{ m} \times 3\text{ m} \times 3\text{ m}$, yielding a total of $N = 1600$ voxels. The TX and RX are centred at $\left[0,-60\text{ m},30\text{ m}\right]^{\text{T}}$ and $\left[0,60\text{ m},30\text{ m}\right]^{\text{T}}$, respectively, with $N_{\text{TX}}=4$ transmitting antennas and $N_{\text{RX}}=4$ receiving antennas. The transmit power is $P_{\text{TX}}=30$ dBm. The center subcarrier frequency is set to $4.9$ GHz. The system employs $T=4$ RISs, each with $50\times50$ elements and with element spacing of $\frac{\lambda}{2} $. These RISs are symmetrically placed at a height of $25$ m and are centred at $\left[\pm d,\pm d,25\text{ m}\right]^{\text{T}}$. Unless otherwise specified, we set $d=30$ m. The amplification factor of ARIS is set to $a=40$ dB \cite{zhou2023framework}. The noise power at the RIS and the RX is set to $\sigma^2_{\text{RIS}}=\sigma^2_{\text{RX}}=-110$ dBm \cite{li2023toward}. The number of signal transmissions is set to $K=50$. The number of non-zero elements in $\mathbf{x}$ is set to $S=10$. The non-zero scattering coefficients follow a Gaussian distribution with a mean of 0.1 and a variance of 0.01.

To evaluate the low-altitude surveillance performance, we employ the following two metrics: 

\textbf{(1) Mean Square Error (MSE):} Measures the per-voxel difference between the predicted $\hat{\mathbf{x}}$ and the ground-truth $\mathbf{x}$:
\begin{equation}
\text{MSE} =\frac{\left \| \mathbf{x}-\hat{\mathbf{x}}\right \|^2_2 }{N}. 
\label{eq32}
\end{equation}

\textbf{(2) Detection Rate (DR):} Assesses the rate of correctly detected non-zero elements in the predicted image. It is defined as the ratio of the number of correctly identified support indices to the total number of non-zero elements in the ground-truth image, given by
\begin{equation}
\text{DR} =\frac{R_\text{d} }{S} ,
\label{eq29}
\end{equation}
where $R_\text{d}$ represents the number of correctly identified support indices.

\subsection{PRIS versus ARIS}
\label{s5-2}
\subsubsection{Noise Comparison}
\label{s5-2-1}
When estimating the channel $h_{i,k,t,j}$ in the ARIS-aided system, the noise at the RX is given by
\begin{equation}
m_{i,k,t,j}=z^{\text{ari}}_{i,k,t,j}+v_{i,k,t,j},
\label{eq30-1}
\end{equation}
where $v_{i,k,t,j}$ is the received noise at the RX, which exists in both PRIS- and ARIS-aided systems. $z^{\text{ari}}_{i,k,t,j}$ represents the noise caused by the reflection-type amplifier, originating from the $t$-th ARIS and arriving at the RX via the RIS-ROI-RX path, which is given as
\begin{equation}
z^{\text{ari}}_{i,k,t,j}=\mathbf{z}_{k,t}^{\text{T}}\mathbf{\Phi }_{k,t} \mathbf{H}_{\text{s}_{t},\text{v}}\text{diag}\left ( \mathbf{x}  \right )  \mathbf{h}_{\text{v},\text{RX}_{j} }.
\label{eq30-2}
\end{equation}

\begin{figure}[t]
\centering
\includegraphics[width=0.47\textwidth]{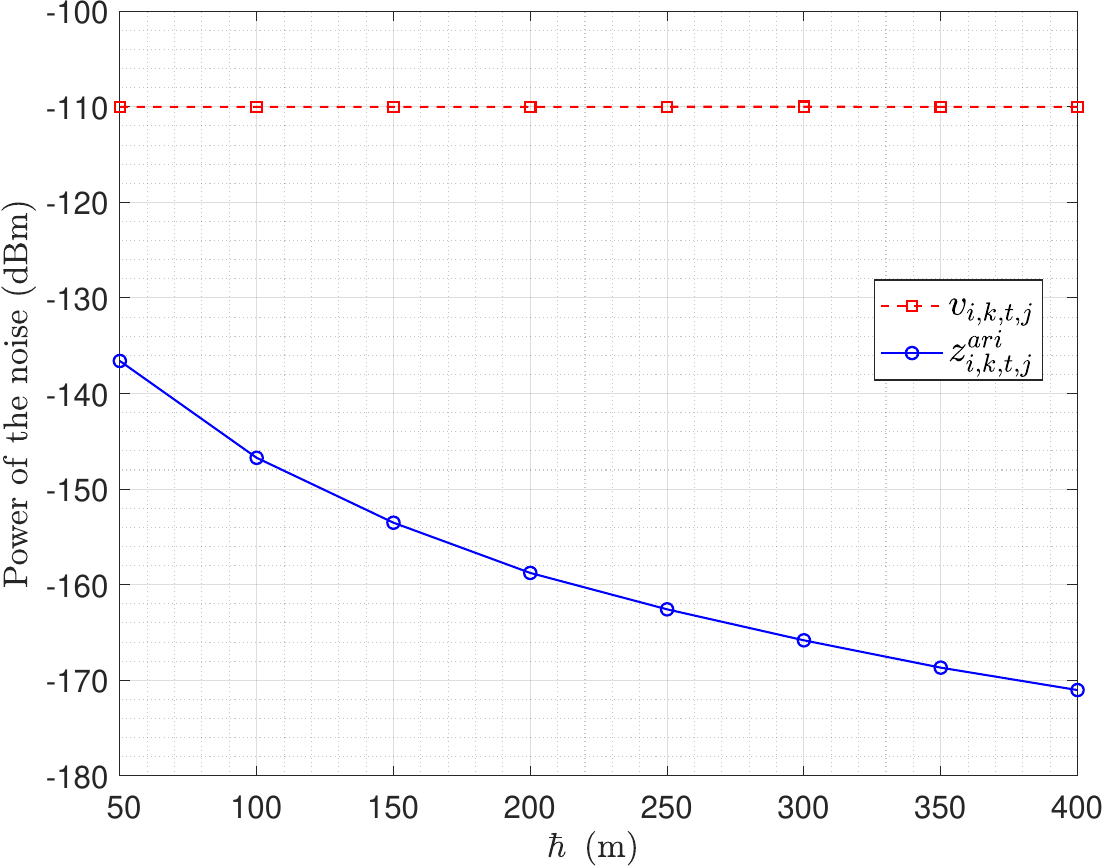}
\caption{Noise comparison between $z^{\text{ari}}_{i,k,t,j}$ and $v_{i,k,t,j}$.}
\label{noise}
\end{figure}

Figure~\ref{noise} presents the comparison between $z^{\text{ari}}_{i,k,t,j}$ and $v_{i,k,t,j}$ at the RX. The result shows that as $h$ increases, the power of $v_{i,k,t,j}$ stabilises at $-110$ dBm. However, the power of $z^{\text{ari}}_{i,k,t,j}$ decreases as $h$ increases and is much lower than $v_{i,k,t,j}$. Consequently, although ARISs introduce and amplify $z^{\text{ari}}_{i,k,t,j}$, they have an extremely small impact on the signal-to-noise ratio of the measurements. 

\subsubsection{Performance Comparison \MakeLowercase{with} the Same Total Power}
\label{s5-2-2}

\begin{figure}[t]
\centering
\subfloat[MSE]{\includegraphics[width=0.47\textwidth]{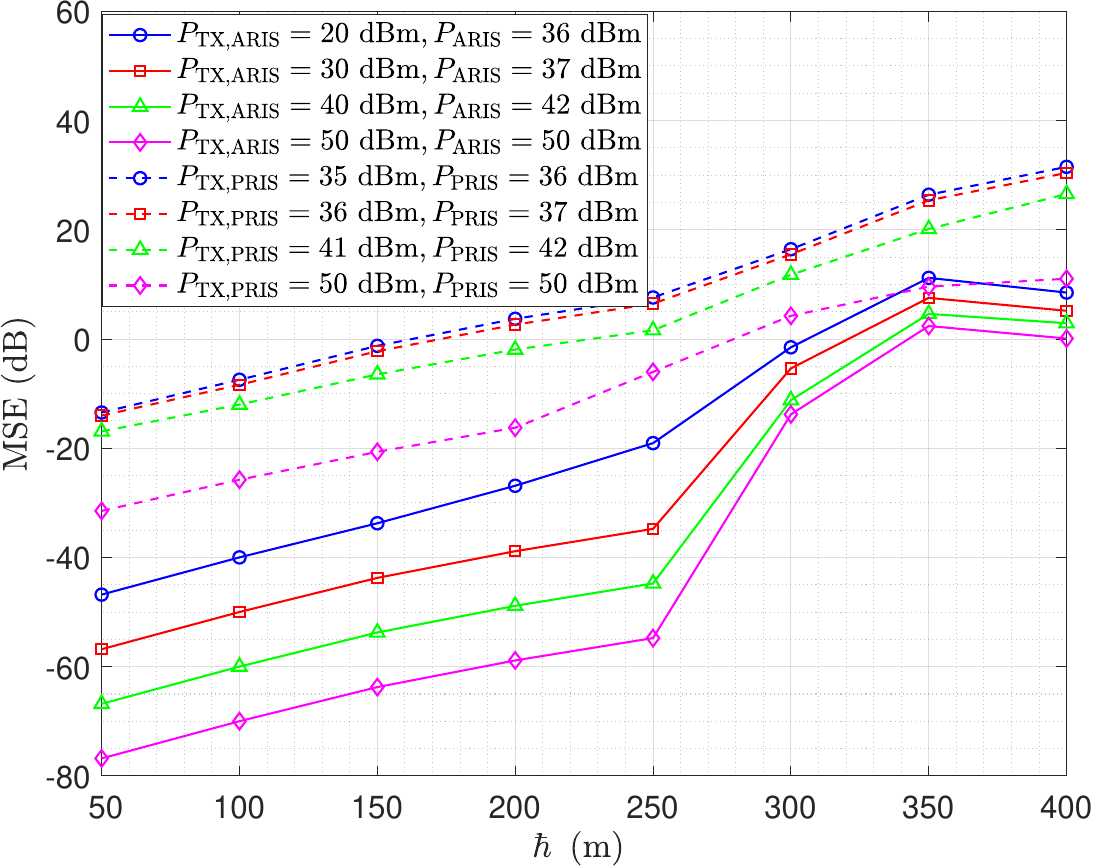}\label{PRISvsARIS}}\\
\subfloat[DR]{\includegraphics[width=0.47\textwidth]{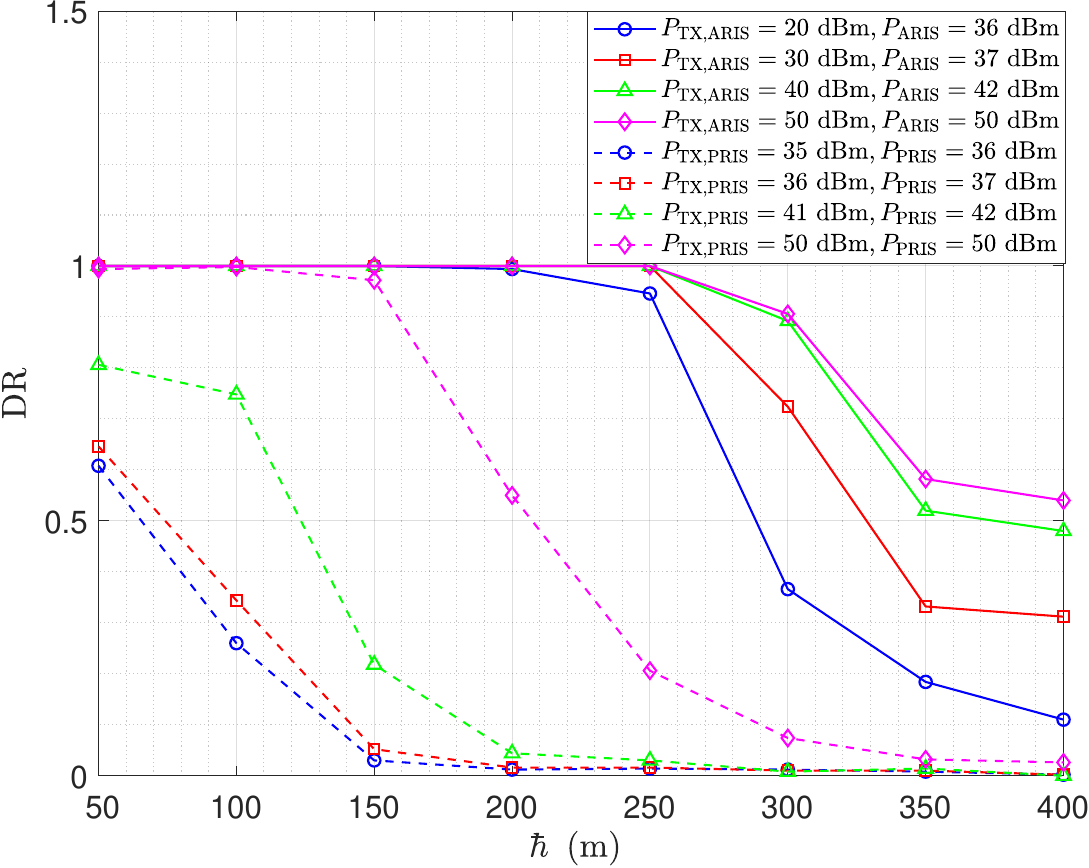}\label{Detection_Rate}}
\caption{ARIS versus PRIS.}\label{fig_a_p}
\end{figure}

For the ARIS-aided system, we set the transmit power $P_{\text{TX,ARIS}}$ to various values and compute the total power $P_{\text{ARIS}}=P_{\text{TX,ARIS}}+\sum_{t=1}^T \left [ P_{\text{M}_t}+ M_t \left( P_{\text{c}}+P_{\text{DC}}\right) \right ] $. For the PRIS-aided system, we set the transmit power $P_{\text{TX,PRIS}}$ to suitable values such that $P_{\text{PRIS}}=P_{\text{ARIS}}$. According to \cite{zhou2023framework}, we set $P_{\text{c}}=-10$ dBm and $P_{\text{DC}}=-5$ dBm. We then evaluate and compare the imaging performance of the two systems under the same total power, with results shown in Figure~\ref{fig_a_p}.

With the same total power, the primary excess power of $P_{\text{TX,PRIS}}$ over $P_{\text{TX,ARIS}}$ is $\sum_{t=1}^{T}M_t P_{\text{DC}}$, which remains constant and independent of $P_{\text{TX,ARIS}}$. Consequently, as $P_{\text{TX,ARIS}}$ increases, the relative power difference between $P_{\text{TX,PRIS}}$ and $P_{\text{TX,ARIS}}$ decreases. For the four cases presented in Figure~\ref{fig_a_p}, $P_{\text{TX,PRIS}}$ is at most 15 dB higher than $P_{\text{TX,ARIS}}$, which is significantly smaller than the amplification factor $a=40$ dB of ARIS. This indicates that the signal amplification of the ARIS-aided system compensates for its lower transmit power of the TX compared to the PRIS-aided system.

As shown in Figure~\ref{PRISvsARIS}, the MSE values for both PRIS- and ARIS-aided systems increase with $\hbar$. At $\hbar = 300$ m, the MSE values increase substantially, suggesting that imaging performance degrades above this height. When $\hbar<300$ m, ARIS consistently outperforms PRIS across all power levels. 

Similarly, Figure~\ref{Detection_Rate} shows that ARIS consistently outperforms PRIS in DR. The DR values for ARIS experience a sharp decline when $\hbar=300$ m, whereas the DR values for PRIS deteriorate significantly when $\hbar>150$ m. We can conclude that the PRIS-aided system hardly achieves effective imaging at $\hbar>150$ m, but the ARIS-aided system still achieves effective imaging results. 

To summarize, although the PRIS-aided system operates with higher transmit power at the BS, the ARIS-aided system still achieves better sensing performance by amplifying RIS-reflected signals.

\subsubsection{Comparison \MakeLowercase{of} Imaging Accuracy \MakeLowercase{per} Watt}
\label{s5-2-3}

\begin{figure}[t]
\centering
\includegraphics[width=0.47\textwidth]{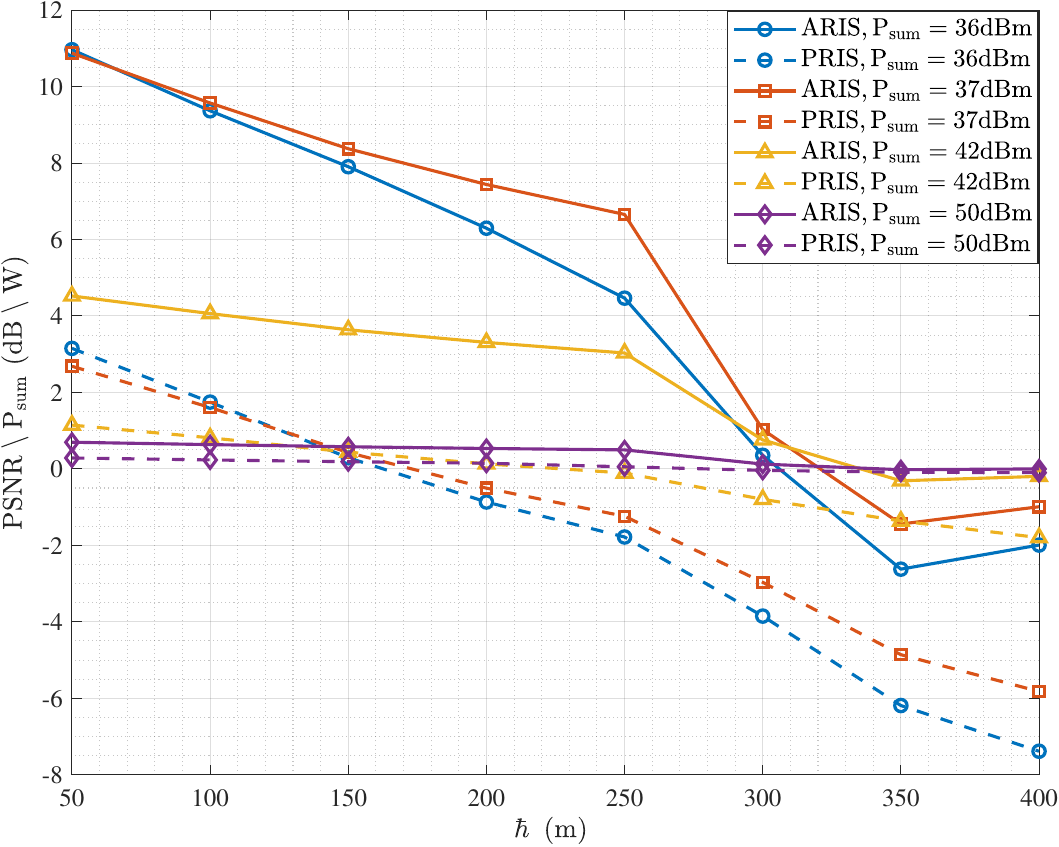}
\caption{PSNR per Watt versus $\hbar$ at various transmit powers.}
\label{PSNR}
\end{figure}

In order to highlight ARIS's benefits and further analyze the power efficiency, we introduce peak signal to noise ratio (PSNR) per Watt, which can be expressed as
\begin{equation}
\frac{\text{PSNR}}{P_{\text{sum}}}=\frac{10\log_{10}{\frac{\text{MAX}^2}{\text{MSE}}}}{P_{\text{sum}}} ,
\label{eq30.5}
\end{equation}
where PSNR is used as a measure of fidelity that is independent of the dynamics of images, and $\text{MAX}$ is the maximum value of $\left | x_n \right | $. $P_{\text{sum}}$ is the total power consumption in the system, which is equal to $P_{\text{PRIS}}$ in the PRIS-aided systems and equal to $P_{\text{ARIS}}$ in the ARIS-aided systems. Figure~\ref{PSNR} shows the value of $\frac{\text{PSNR}}{P_{\text{sum}}}$ in PRIS- and ARIS-aided systems. Under different power levels, the value of $\frac{\text{PSNR}}{P_{\text{sum}}}$ in ARIS-aided systems is consistently higher than that in PRIS-aided systems, and this advantage is more pronounced when $P_{\text{sum}}$ is low. The value of $\frac{\text{PSNR}}{P_{\text{sum}}}$ decreases as the height increases. In the ARIS-aided systems, with the increase in $P_{\text{sum}}$, the value of $\frac{\text{PSNR}}{P_{\text{sum}}}$ improves when $P_{\text{sum}}$ is small. However, when $P_{\text{sum}}$ is large, the PSNR improvement becomes insignificant, leading to a decrease in $\frac{\text{PSNR}}{P_{\text{sum}}}$.

\subsection{Limit of ROI height for ARIS-aided systems}
\label{s5-3}

\begin{figure}[t]
\centering
\includegraphics[width=0.47\textwidth]{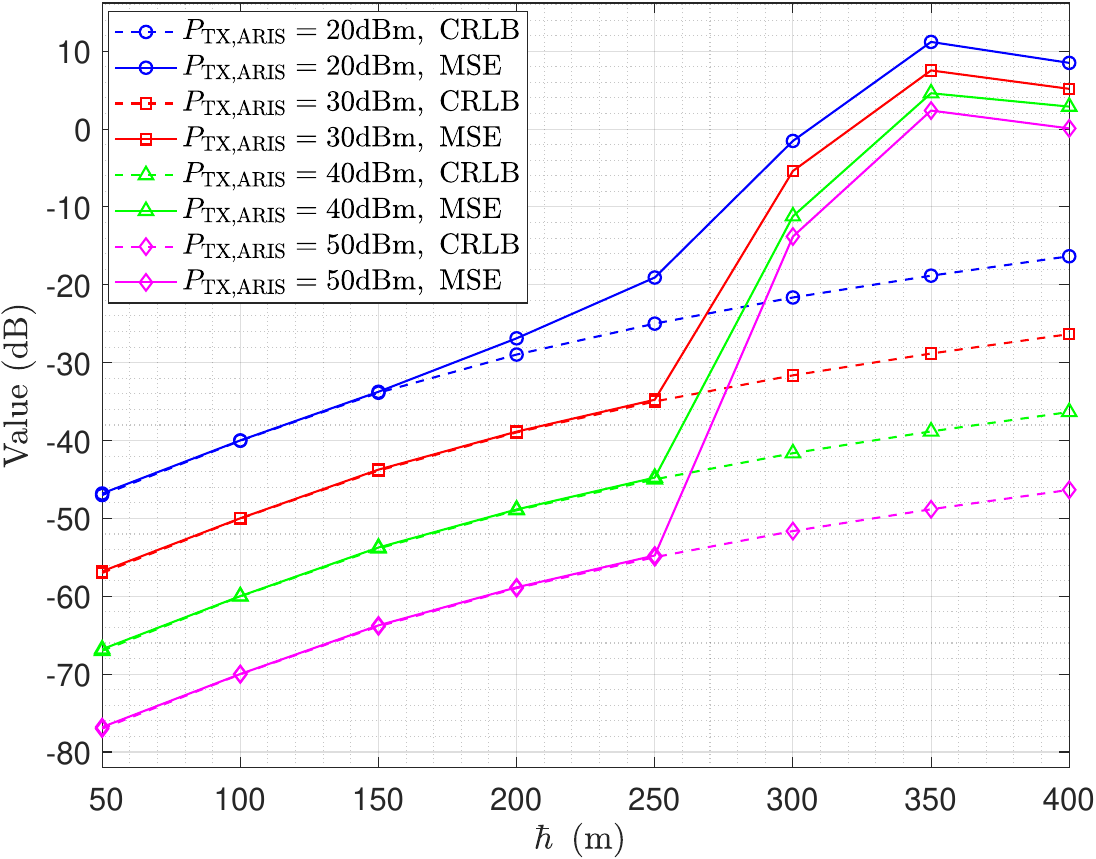}
\caption{MSE and CRLB versus $\hbar$ at various transmit power.}
\label{fig_height_limit}
\end{figure}

To establish the limit of ROI height for ARIS-aided systems, we study the relationship between MSE, CRLB, and the height of the ROI across different transmit powers, as presented in Figure~\ref{fig_height_limit}. The limit of imaging height can be defined by two criteria: first, the height where the MSE value exceeds -20 dB, indicating significant sensing performance degradation; second, the height where the MSE value begins to deviate substantially from the CRLB value. The CRLB value can be obtained according to EQ.~\eqref{eq31}.

When $\hbar$ ranges from $50$ to $250\text{ m}$, the MSE values increase slowly and are highly close to the CRLB values. At the ROI height of $300\text{ m}$, the MSE values exceed the threshold of $-20\ \text{dB}$ for various transmit power. Simultaneously, the gap between the MSE values and the CRLB values initiates at $250\text{ m}$. Consequently, we can conclude that $250-300\text{ m}$ is the limit of ROI height for ARIS-aided systems.

\subsection{Influence of other system parameters}
\label{s5-4}

In this subsection, we verify the influences of various system parameters on the sensing performance using MSE. Moreover, we compare the MSE trends with the conclusions drawn from the CRLB in Sec.~\ref{s4-2}.

\begin{figure}[t]
\centering
\includegraphics[width=0.47\textwidth]{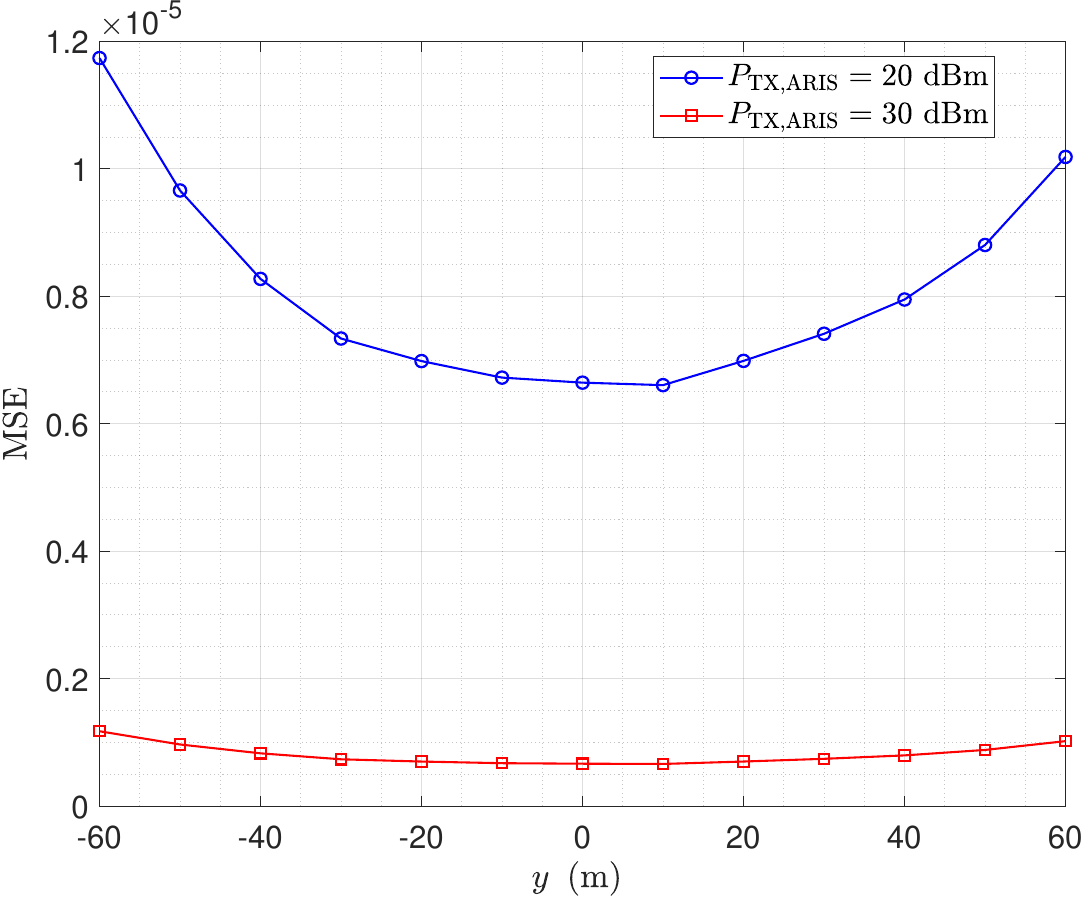}
\caption{MSE versus the RX position.}
\label{fig:NMSE-RX}
\end{figure}

Figure~\ref{fig:NMSE-RX} illustrates the MSE values, where the y-axis of the RX ranges from $-60\text{ m}$ to $60\text{ m}$, while maintaining a constant height of $30 \text{ m}$ and a constant x-axis of $0$. The MSE reaches a distinct minimum at a specific RX location, where $0 \le y \le 10 \text{ m}$. It corresponds to the optimal RX position derived in Sec.~\ref{s4-2}. In Sec.~\ref{s4-2}, we have also concluded that the distances between the RX and the voxels, and the TX-RIS-ROI path, jointly influence the CRLB values. They influence the MSE values in a similar way, as shown in Figure~\ref{fig:NMSE-RX}. Consequently, the optimal y-axis of the RX position is determined to be in the range of $\left [  0, 10\text{ m}  \right ] $.

\begin{figure}[t]
\centering
\includegraphics[width=0.47\textwidth]{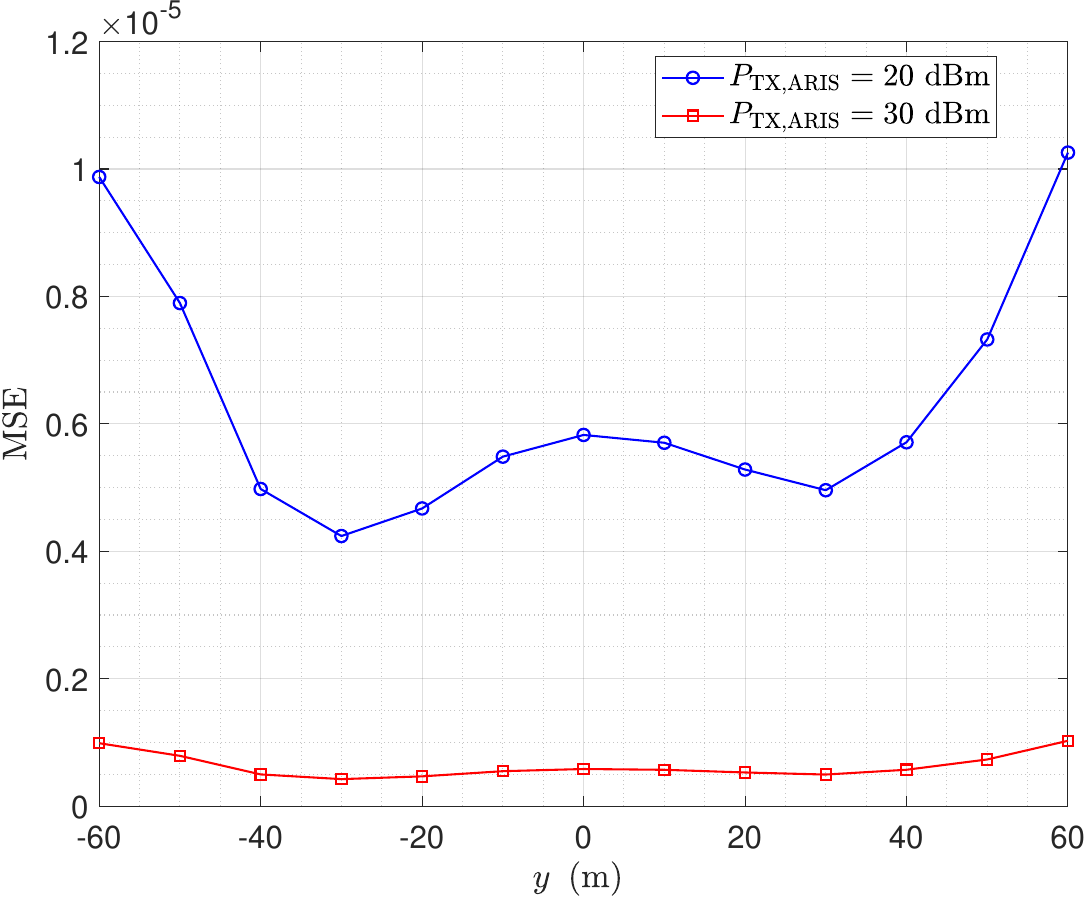}
\caption{MSE versus the TX position.}
\label{fig:nmse-TX}
\end{figure}

Figure~\ref{fig:nmse-TX} illustrates the MSE values, where the y-axis of the TX ranges from $-60\text{ m}$ to $60\text{ m}$, while maintaining a constant height of $30\text{ m}$ and a constant x-axis value of $30\, \text{ m}$. Consistent with the analysis in Sec.~\ref{s4-2},  the MSE reaches local minima when the TX is positioned directly above the center of any RIS. This further validates the theoretical insights from Sec.\ref{s4-2}. 

\begin{figure}[t]
\centering
\includegraphics[width=0.47\textwidth]{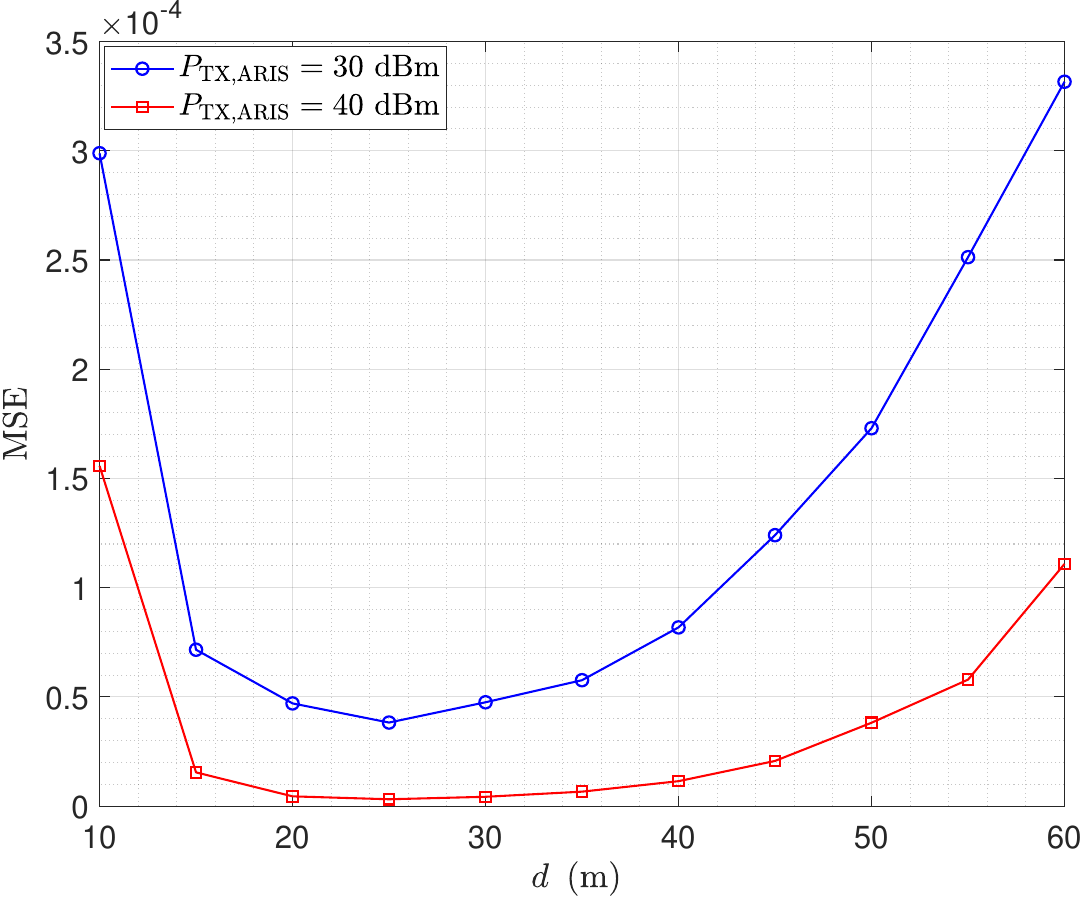}
\caption{MSE versus the RIS positions.}
\label{fig:nmse-RIS}
\end{figure}

Figure~\ref{fig:nmse-RIS} illustrates the MSE values, where $d$ ranges from $10\text{ m}$ to $60\text{ m}$. The MSE value shows a trend of decreasing followed by increasing, with a local minimum occurring at $d = 25 \text{ m}$. Notably, Figure~\ref{fig:nmse-RIS} differs from the one presented in Figure~\ref{fig9}, as the distance $d$ influences MSE in two distinct ways. On the one hand, the path length varies with $d$, which primarily affects the performance as observed in Figure~\ref{fig9}. On the other hand, a larger imaging aperture, which is positively correlated with $d$, results in a lower MSE, further influencing the system's performance. Consequently, as $d$ increases, these two factors, counterbalancing each other, reach an optimal trade-off at $d = 25 \text{ m}$, where the MSE is minimized.

\section{Conclusion}{}
\label{CONCLUSION}

This study proposes a RIS-aided cooperative ISAC network for surveillance in low-altitude airspace, which addresses the limitations of high deployment costs and weak signals in conventional methods. 
The low-altitude surveillance is modeled as an imaging problem based on CS theory, which can be solved through the SP algorithm. 
We derive the CRLB of the proposed RIS-aided low-altitude imaging system, and analyze the impacts of various system parameters on sensing performance. 
Through analysis and simulations, we derive three key findings. First, ARIS provides superior performance over PRIS under the same power constraints through amplifying signals. Second, ARIS-aided systems can achieve surveillance at low altitudes below 300 meters. Third, our work provides insights into the system parameter design: the RX should be positioned near the ROI center, the TX should be close to one of the RISs, and RIS spacing should realize a balance between path length and imaging aperture.
In conclusion, the proposed RIS-aided cooperative ISAC network has the potential to significantly improve the efficiency and accuracy of low-altitude surveillance, offering a promising solution for future network deployments. Future work will focus on the optimization of RIS deployment and the handling of more complex non-ideal interference.

\bibliographystyle{IEEEtran}
\bibliography{myref}

\begin{thebibliography}{10}
\providecommand{\url}[1]{#1}
\csname url@samestyle\endcsname
\providecommand{\newblock}{\relax}
\providecommand{\bibinfo}[2]{#2}
\providecommand{\BIBentrySTDinterwordspacing}{\spaceskip=0pt\relax}
\providecommand{\BIBentryALTinterwordstretchfactor}{4}
\providecommand{\BIBentryALTinterwordspacing}{\spaceskip=\fontdimen2\font plus
\BIBentryALTinterwordstretchfactor\fontdimen3\font minus
  \fontdimen4\font\relax}
\providecommand{\BIBforeignlanguage}[2]{{%
\expandafter\ifx\csname l@#1\endcsname\relax
\typeout{** WARNING: IEEEtran.bst: No hyphenation pattern has been}%
\typeout{** loaded for the language `#1'. Using the pattern for}%
\typeout{** the default language instead.}%
\else
\language=\csname l@#1\endcsname
\fi
#2}}
\providecommand{\BIBdecl}{\relax}
\BIBdecl

\bibitem{luo2025toward}
Y.~Wang, G.~Sun, Z.~Sun, J.~Wang, J.~Li, C.~Zhao, J.~Wu, S.~Liang, M.~Yin,
  P.~Wang, D.~Niyato, S.~Sun, and D.~In~Kim, ``Toward realization of
  low-altitude economy networks: Core architecture, integrated technologies,
  and future directions,'' \emph{IEEE Trans. Cognit. Commun. Networking},
  vol.~11, no.~5, pp. 2788--2820, Aug. 2025.

\bibitem{how_to_achieve}
M.~Liu, H.~Xiao, G.~Pan, J.~Zhou, W.~Li, and X.~Xi, ``How to achieve
  large-scale development in the low-altitude economy,'' in \emph{Proc. Int.
  Symp. Microwave Antenna Propag. EMC Technol. Wireless Commun. {(MAPE)}}, Nov.
  2024, pp. 1--4.

\bibitem{11059622}
M.~Wei, Y.~Zhao, and J.~Yu, ``Research on low altitude network coverage
  solutions,'' in \emph{Proc. Int. Wireless Commun. Mob. Comput. (IWCMC)}, May
  2025, pp. 412--417.

\bibitem{9631203}
IEEE, ``{IEEE} standard for a framework for structuring low-altitude airspace
  for unmanned aerial vehicle ({UAV}) operations,'' \emph{IEEE Std
  1939.1-2021}, pp. 1--94, Dec. 2021.

\bibitem{khan2022detection}
M.~A. Khan, H.~Menouar, A.~Eldeeb, A.~Abu-Dayya, and F.~D. Salim, ``On the
  detection of unauthorized drones—techniques and future perspectives: A
  review,'' \emph{IEEE Sens. J.}, vol.~22, no.~12, pp. 11\,439--11\,455, Jun.
  2022.

\bibitem{anti-drone}
X.~Shi, C.~Yang, W.~Xie, C.~Liang, Z.~Shi, and J.~Chen, ``Anti-drone system
  with multiple surveillance technologies: Architecture, implementation, and
  challenges,'' \emph{IEEE Commun. Mag.}, vol.~56, no.~4, pp. 68--74, Apr.
  2018.

\bibitem{wu2024vehicle}
G.~Wu, F.~Zhou, K.~K. Wong, and X.-Y. Li, ``A vehicle-mounted radar-vision
  system for precisely positioning clustering {UAV}s,'' \emph{IEEE J. Sel.
  Areas Commun.}, vol.~42, no.~10, pp. 2688--2703, Oct. 2024.

\bibitem{learned_off_grid}
Y.~Huang, J.~Yang, S.~Xia, C.-K. Wen, and S.~Jin, ``Learned off-grid imager for
  low-altitude economy with cooperative {ISAC} network,'' \emph{IEEE Trans.
  Wireless Commun.}, pp. 1--1, Sep. 2025.

\bibitem{li2023integrated}
H.~Li, J.~Xu, C.~Sun, S.~Wang, X.~Wang, and H.~Zhang, ``Integrated sensing and
  communication: {3GPP} standardization progress,'' in \emph{Proc. Int. Symp.
  Model. Optim Mobile Ad Hoc Wireless Networks (WiOpt)}, Aug. 2023, pp. 1--7.

\bibitem{li2023toward}
R.~Li, Z.~Xiao, and Y.~Zeng, ``Toward seamless sensing coverage for cellular
  multi-static integrated sensing and communication,'' \emph{IEEE Trans.
  Wireless Commun.}, vol.~23, no.~6, pp. 5363--5376, Jun. 2024.

\bibitem{wang2024heterogeneous}
Z.~Wang and V.~Wong, ``Heterogeneous graph neural network for cooperative
  {ISAC} beamforming in cell-free {MIMO} systems,'' in \emph{Proc. Annu. Int.
  Conf. Mobile Comput. Networking}, Dec. 2024, pp. 2161--2172.

\bibitem{he2024device}
J.~He, C.~Vanwynsberghe, H.~Chen, C.~Huang, and A.~Fakhreddine, ``Device-free
  {3D} drone localization in {RIS}-assisted mmwave {MIMO} networks,'' in
  \emph{Proc. GLOBECOM IEEE Global Commun. Conf.}, Dec. 2024, pp. 4436--4441.

\bibitem{11008547}
G.~Mylonopoulos, L.~Venturino, E.~Grossi, S.~Buzzi, and C.~D’Elia,
  ``Integrated communication and {RIS}-aided track-before-detect radar
  sensing,'' \emph{IEEE Open J. Commun. Soc.}, vol.~6, pp. 4519--4532, May
  2025.

\bibitem{wang2024reconfigurable}
J.~Wang, W.~Tang, J.~C. Liang, L.~Zhang, J.~Y. Dai, X.~Li, S.~Jin, Q.~Cheng,
  and T.~J. Cui, ``Reconfigurable intelligent surface: Power consumption
  modeling and practical measurement validation,'' \emph{IEEE Trans. Commun.},
  vol.~72, no.~9, pp. 5720--5734, Sep. 2024.

\bibitem{encinas2025riloco}
G.~Encinas-Lago, V.~Sciancalepore, H.~Wymeersch, M.~Di~Renzo, and
  X.~Costa-Pérez, ``{RiLoCo}: An {ISAC}-oriented {AI} solution to build
  {RIS}-empowered networks,'' \emph{IEEE Trans. Wireless Commun.}, vol.~24,
  no.~10, pp. 8221--8235, May 2025.

\bibitem{huang2024ris}
Y.~Huang, J.~Yang, C.-K. Wen, and S.~Jin, ``{RIS}-aided single-frequency {3D}
  imaging by exploiting multi-view image correlations,'' \emph{IEEE Trans.
  Commun.}, vol.~72, no.~8, pp. 5003--5018, Aug. 2024.

\bibitem{huang2023joint}
Y.~Huang, J.~Yang, W.~Tang, C.-K. Wen, S.~Xia, and S.~Jin, ``Joint localization
  and environment sensing by harnessing {NLOS} components in {RIS}-aided mmwave
  communication systems,'' \emph{IEEE Trans. Wireless Commun.}, vol.~22,
  no.~12, pp. 8797--8813, Dec. 2023.

\bibitem{zhu2023ris}
Y.~Liu and W.~Yu, ``{RIS}-assisted joint sensing and communications via
  fractionally constrained fractional programming,'' in \emph{GLOBECOM IEEE
  Global Commun. Conf.}, Dec. 2024, pp. 650--655.

\bibitem{li2024radio}
Z.~Li, A.~Dubey, S.~Shen, N.~K. Kundu, J.~Rao, and R.~Murch, ``Radio
  tomographic imaging with reconfigurable intelligent surfaces,'' \emph{IEEE
  Trans. Wireless Commun.}, vol.~23, no.~11, pp. 15\,784--15\,797, Nov. 2024.

\bibitem{zhou2023framework}
G.~Zhou, C.~Pan, H.~Ren, D.~Xu, Z.~Zhang, J.~Wang, and R.~Schober, ``A
  framework for transmission design for active {RIS}-aided communication with
  partial {CSI},'' \emph{IEEE Trans. Wireless Commun.}, vol.~23, no.~1, pp.
  305--320, Jan. 2023.

\bibitem{saikia2024ris}
P.~Saikia, A.~Jee, K.~Singh, C.~Pan, W.-J. Huang, and T.~A. Tsiftsis,
  ``{RIS}-aided integrated sensing and communication systems: {STAR-RIS} vs
  passive {RIS}?'' \emph{IEEE Open J. Commun. Soc.}, vol.~5, pp. 7954--7973,
  Dec. 2024.

\bibitem{khoshafa2021active}
M.~H. Khoshafa, T.~M. Ngatched, M.~H. Ahmed, and A.~R. Ndjiongue, ``Active
  reconfigurable intelligent surfaces-aided wireless communication system,''
  \emph{IEEE Commun. Lett.}, vol.~25, no.~11, pp. 3699--3703, Nov. 2021.

\bibitem{zhang2025research}
J.~Zhang, F.~Lin, X.~Ji, and H.~Sun, ``Research on total power optimization of
  {NOMA-ISAC} system assisted by dual active {RIS},'' in \emph{Proc. Int. Conf.
  Sens. Inf. Technol.}, Mar. 2025, pp. 141--145.

\bibitem{zhang2022active}
Z.~Zhang, L.~Dai, X.~Chen, C.~Liu, F.~Yang, R.~Schober, and H.~V. Poor,
  ``Active {RIS} vs. passive {RIS}: Which will prevail in {6G}?'' \emph{IEEE
  Trans. Commun.}, vol.~71, no.~3, pp. 1707--1725, Mar. 2022.

\bibitem{huang2024fourier}
Y.~Huang, J.~Yang, W.~Tang, C.-K. Wen, and S.~Jin, ``Fourier transform-based
  wavenumber domain {3D} imaging in {RIS}-aided communication systems,''
  \emph{IEEE Trans. Wireless Commun.}, vol.~23, no.~10, pp. 13\,872--13\,888,
  Oct. 2024.

\bibitem{chen2025experimental}
F.-J. Chen, C.-K. Wen, and D.-M. Chian, ``Experimental evaluation of multiple
  active {RIS}s for {5G} {MIMO} commercial networks,'' \emph{IEEE Wireless
  Commun. Lett.}, vol.~14, no.~8, pp. 2511--2515, May 2025.

\bibitem{wang2025reconfigurable}
X.~Wang, Y.~Huang, J.~Yang, Y.~Han, and S.~Jin, ``Reconfigurable intelligent
  surface aided integrated communication and localization with a single access
  point,'' \emph{arXiv preprint arXiv:2505.02444}, May 2025.

\bibitem{dai2009subspace}
W.~Dai and O.~Milenkovic, ``Subspace pursuit for compressive sensing signal
  reconstruction,'' \emph{IEEE Trans. Inf. Theory}, vol.~55, no.~5, pp.
  2230--2249, May 2009.

\bibitem{10621011}
C.~Ruan, Z.~Zhang, H.~Jiang, Y.~Qi, J.~Dang, and L.~Wu, ``Low complexity
  orthogonal matching pursuit-based near-field channel estimation in {XL-MIMO}
  systems,'' \emph{IEEE Commun. Lett.}, vol.~28, no.~12, pp. 2859--2863, Aug.
  2024.

\end{thebibliography}

\end{document}